\documentclass[%
reprint,
nofootinbib,
 amsmath,amssymb,
prd,
floatfix,
]{revtex4-2}

\usepackage{graphicx}% Include figure files
\usepackage{dcolumn}% Align table columns on decimal point
\usepackage{bm}% bold math
\usepackage{hyperref}% add hypertext capabilities
\usepackage{physics}
\usepackage{slashed}
\usepackage{floatrow}
\usepackage{subfig}
\usepackage[tableposition=top]{caption}
\usepackage{times,txfonts}

\newcommand{\e}{\mathrm{e}}
\newcommand{\iu}{\mathrm{i}}
\newcommand{\g}[1]{\gamma^{#1}}

\newcommand{\us}[4]{u_{#1,#2}^{\left(#3\right)}\left(#4\right)}
\newcommand{\ubs}[4]{\Bar{u}_{#1,#2}^{\left(#3\right)}\left(#4\right)}
\newcommand{\vs}[4]{v_{#1,#2}^{\left(#3\right)}\left(#4\right)}
\newcommand{\vbs}[4]{\Bar{v}_{#1,#2}^{\left(#3\right)}\left(#4\right)}

\newcommand{\lb}{\lambda_B^2}

\newcommand{\Ia}[3]{\mathcal{I}^{#1,#2}\left(#3\right)}

\renewcommand{\Vec}[1]{\ensuremath{\bm{#1}}}
\renewcommand{\vec}[1]{\ensuremath{\bm{#1}}}

\newcommand{\f}[2]{f^{#1}\left(#2\right)}

\newcommand{\CC}[2]{C_{#1}^{\mathrm{ST}}\left(C_{#2'}^{\mathrm{ST}}\right)^*}
\newcommand{\m}[4]{\mathcal{M}^{(#1,#2)}_{#3,#4}}

\newcommand{\Ik}[2]{I_{#1,#2}\left(\frac{k_\perp^2\lb}{2}\right)}
\newcommand{\Ikp}[2]{I_{#1,#2}\left(\frac{\left(k'_\perp\right)^2\lb}{2}\right)}

\DeclareMathOperator{\sign}{sign}

\newcommand\Tstrut{\rule{0pt}{1.6\normalbaselineskip}}         % = `top' strut
\newcommand\Bstrut{\rule[-1.3\normalbaselineskip]{0pt}{0pt}}   % = `bottom' strut

\begin{document}

\title{Evaluation of QED cross sections in strong magnetic fields}

\author{Olavi Kiuru}
\email{olavi.kiuru@helsinki.fi}
\affiliation{%
Department of Physics, P.O.~Box 64, FI-00014 University of Helsinki, Finland
}%

\date{\today}

\begin{abstract}
Quantum electrodynamics (QED) becomes nonlinear when the magnetic field strength surpasses the critical Schwinger limit $B_\mathrm{Q} \approx 4.41\cdot 10^{13}\;$G. This limit is surpassed, for example, in the magnetospheres of a specific class of neutron stars known as magnetars, which has important consequences for magnetospheric plasma dynamics due to modifications in scattering cross sections.
Using a formalism previously applied to the study of magnetic catalysis, I calculate the cross sections of all tree-level 1-to-2, 2-to-1, and 2-to-2 particle QED scattering processes that do not include a photon propagator. The calculations are done in a strong background magnetic field and the results are implemented into an open-source Python package. This article focuses on presenting the formalism and computational techniques required for the calculations, while the impact of the results on, e.g., magnetospheric plasma dynamics is discussed in a companion letter \citep{kiuru_qed_2026}.
\end{abstract}

\maketitle

\section{\label{sec:intro}Introduction}
The question of how quantum electrodynamics (QED) becomes modified in the presence of an electromagnetic background field was first addressed by \citet{heisenberg_folgerungen_1936}, who derived the Euler-Heisenberg Lagrangian
\begin{equation}\label{eq:EHLagrangian}
\begin{split}
    &\mathcal{L}_\mathrm{EH} = -\frac{1}{4}\mathcal{F}^2-\frac{1}{8\pi^2}\int_0^\infty\frac{\dd{s}}{s^3}\e^{-m_e^2 s} \\
    &\cross\left[\frac{(es)^2}{4}\frac{\Re \cosh(es\sqrt{\frac{\mathcal{F}^2+\iu \mathcal{F} \vdot \mathcal{G}}{2}})}{\Im\cosh(es\sqrt{\frac{\mathcal{F}^2+\iu \mathcal{F} \vdot \mathcal{G}}{2}})}\mathcal{F}\vdot \mathcal{G}-\frac{(es)^2}{6}\mathcal{F}^2-1\right].
\end{split}
\end{equation}
Here, $\mathcal{F_{\mu\nu}}$ is the electromagnetic tensor of the background field and $\mathcal{G_{\mu\nu}}$ is its Hodge dual. The weak-field limit of the Euler-Heisenberg Lagrangian
\begin{equation}
    \mathcal{L}_\mathrm{EH}^{\mathrm{wf}}= \frac{1}{2}\left(\vec{E}^2-\vec{B}^2\right)+\frac{2\alpha_e^2}{45m_e^4}\left[\left(\vec{E}^2-\vec{B}^2\right)^2+7\left(\vec{E}\vdot\vec{B}\right)^2\right]
\end{equation}
shows explicitly that a background field gives rise to nonlinear terms in Maxwell's equations. These nonlinearities must be taken into account when the field strength approaches the Schwinger limit $B_\mathrm{Q}=m_e^2/e\approx 4.41\cdot 10^{13}\,$G or ${E_\mathrm{Q}=m_e^2/e\approx1.32\cdot 10^{18}\;}$V/m (with $c=\hbar=1$). Nowadays, the Euler-Heisenberg Lagrangian is often discussed in the context of the Schwinger proper time method \citep{schwinger_gauge_1951} that turns the problem of finding the Lagrangian into a problem that resembles the calculation of the time-evolution of a quantum state with the proper time variable $s$.

In this article, I will focus on the special case of a constant magnetic field and no electric field. This is a good approximation of the electromagnetic field in a magnetar magnetosphere, where the curvature radius of the magnetic field lines is much larger than the magnetic length $\lambda_B\equiv \abs{qB}^{-\frac{1}{2}}$ and the electric field is expected to be much weaker than the magnetic field. In a constant magnetic field, the energies of electrons and positrons ($e^\pm$) become quantized into so-called Landau levels \citep{akhiezer_quantum_1965}
\begin{equation}\label{eq:LandauEnergy}
    E^2 = m_e^2 + \left(p^z\right)^2 + 2n\abs{qB},
\end{equation}
where $n\in \mathbb{N}$ is the Landau level index and $q$ is the charge of the fermion ($q=-e$ for an electron). As will be discussed further in Sec.~\ref{sec:setup}, the quantization of the energy levels of $e^\pm$ allows us to perform a Landau level projection of the proper time integral in Eq.~\eqref{eq:EHLagrangian}, where the integral is replaced by a sum over Landau levels.

A solid understanding of QED in the nonlinear regime is vital for the study of magnetars, i.e., strongly magnetized neutron stars, whose magnetic fields can reach $B\approx 30\,B_\mathrm{Q}$ in the magnetosphere. Magnetars were first proposed by \citet{duncan_formation_1992} to explain the origin of observed extreme transient electromagnetic radiation phenomena. The magnetar emission spectra have been observed to have a double-peak structure that currently cannot be derived from first principles \citep{kaspi_magnetars_2017, rea_magnetars_2025}. QED processes like pair creation have been found to be essential for the plasma dynamics of pulsars \citep{goldreich_pulsar_1969, sturrock_model_1971} and similar results have been obtained also for magnetars \citep{beloborodov_corona_2007, zeng_origin_2025, zhang_quantum_2025}. However, current magnetar simulations do not fully incorporate the effects of the extreme magnetic field on QED scattering processes because all cross sections have not been calculated. 

\begin{table}
    \begin{tabular}{l l c c}
    Interaction & Name & Diagram & Refs. \\[0.5ex]
    \hline \hline
     $e^\pm \rightarrow e^\pm + \gamma$ & SR & \raisebox{-0.45\height}{\includegraphics[scale=0.35]{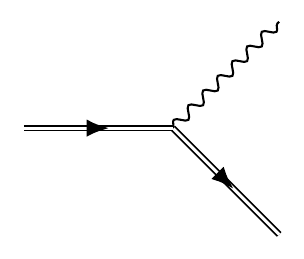}} \Tstrut \Bstrut & \citep{herold_cyclotron_1982}\\
     $\gamma \rightarrow e^+ + e^-$ & One-$\gamma$ PC & \raisebox{-0.45\height}{\includegraphics[scale=0.35]{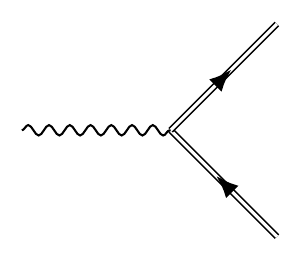}} \Bstrut & \citep{kostenko_qed_2018}\\
     \hline
     $e^+ + e^- \rightarrow \gamma$ & One-$\gamma$ PA & \raisebox{-0.45\height}{\includegraphics[scale=0.35]{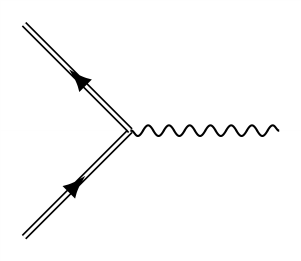}}
     \Tstrut \Bstrut & \citep{wunner_comparison_1979,daugherty_pair_1980}\\
     $e^\pm + \gamma \rightarrow e^\pm$ & SSA & \raisebox{-0.45\height}{\includegraphics[scale=0.35]{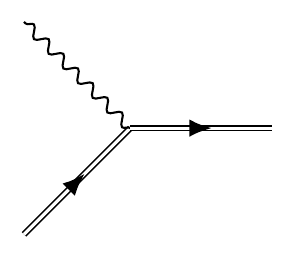}} \Bstrut & \\
     \hline
     $\gamma + \gamma \rightarrow e^- + e^+ $ & Two-$\gamma$ PC & \raisebox{-0.45\height}{\includegraphics[scale=0.35]{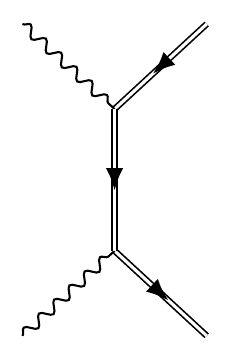}} \Bstrut & \citep{kozlenkov_two-photon_1986,thompson_electrodynamics_2008}\\
     $e^- + e^+ \rightarrow \gamma + \gamma$ & Two-$\gamma$ PA & \raisebox{-0.45\height}{\includegraphics[scale=0.35]{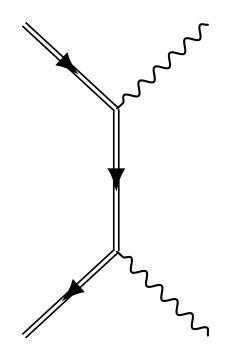}}  \Tstrut \Bstrut & \cite{daugherty_pair_1980, lewicka_two-photon_2013}\\
     $e^\pm + \gamma \rightarrow e^\pm + \gamma$ & Compton & \raisebox{-0.45\height}{\includegraphics[scale=0.35]{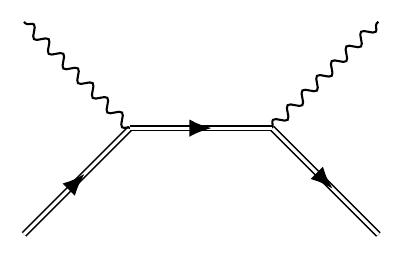}} \Bstrut & \citep{mushtukov_compton_2016, gonthier_compton_2014}\\
     \hline
     $\e^\pm + e^\pm \to e^\pm + e^\pm$ & Møller & \raisebox{-0.45\height}{\includegraphics[scale=0.35]{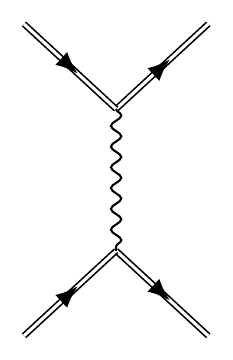}} &  \citep{tiwari_lowest-order_2018}\\
     $\e^+ + e^- \to e^+ + e^-$ & Bhabha & \raisebox{-0.45\height}{\includegraphics[scale=0.35]{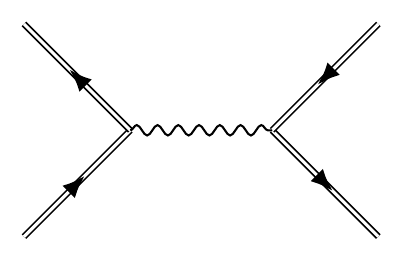}} &  \citep{sis_eellmathmrow_2023,tiwari_lowest-order_2018}\\
    \end{tabular}
\caption{QED scattering processes in magnetized astrophysical plasmas. Electrons are denoted with $e^-$, positrons with $e^+$, and photons with $\gamma$. The following abbreviations are used: synchrotron radiation (SR), pair creation (PC), pair annihilation (PA), and synchrotron self-absorption (SSA).}\label{table:QEDProcesses}
\end{table}

Due to their relevance to high-energy astrophysics, QED processes in strong background magnetic fields have been studied extensively \citep{adler_photon_1970, wunner_comparison_1979, herold_cyclotron_1982, daugherty_pair_1980, daugherty_compton_1986, gonthier_compton_2000, gonthier_compton_2014, mushtukov_compton_2016, thompson_electrodynamics_2008, lewicka_two-photon_2013, kostenko_qed_2018, kostenko_qed_2019, harding_physics_2006, harding_physics_2020, melrose_quantum_2013, kozlenkov_two-photon_1986}. However, the existing studies focus either on one specific process (see, e.g., \citep{mushtukov_compton_2016}) or make assumptions about the occupied Landau levels that limit the applicability of their result (see, e.g., \citep{kostenko_qed_2018, kostenko_qed_2019}). In this paper, I adopt a new formalism for calculating the cross sections of QED scattering processes in a strong background magnetic field, originally used in the context of magnetic catalysis \citep{shovkovy_magnetic_2013, miransky_quantum_2015}. The calculated cross sections are valid for all three regions of interest: $b\equiv B/B_\mathrm{Q} \ll 1$, $b\approx 1$, and $b \gg 1$. Table~\ref{table:QEDProcesses} lists the main tree-level processes that are expected to contribute to the plasma dynamics of magnetar magnetospheres.

The study of strong-field QED (SFQED) is not only important for applications in high-energy astrophysics, but is also of interest for laser plasma physics (for recent reviews see, e.g., \citep{fedotov_advances_2023, gonoskov_charged_2022}). In laser plasma physics, the background field is an electromagnetic plane wave, while astrophysical applications often use a constant magnetic field. Regardless of this difference, advances in the understanding of SFQED are of use for both disciplines. 

This article focuses on the technical details of calculating QED cross sections in a strong magnetic field, while the impact and physical context of the results is discussed in a companion letter \citep{kiuru_qed_2026}. The structure of this article is the following. In Sec.~\ref{sec:setup}, I will introduce the new formalism and discuss its validity. Subsequently, the $\mathcal{O}(\alpha_e)$ and $\mathcal{O}(\alpha_e^2)$ QED scattering cross sections are discussed in Sec.~\ref{sec:3part} and Sec.~\ref{sec:4part}, respectively. I will also briefly discuss future applications of the presented formalism in Sec.~\ref{sec:otherAlphaTwo} and Sec.~\ref{sec:higherOrder}.

\section{\label{sec:setup}Setup}

The calculations in this article are done in an environment with a constant homogeneous magnetic field pointing in the $z$-direction
\begin{equation}
    \vec{B} = (0,0,B)
\end{equation}
and no electric field. In a magnetar magnetosphere, both the rotational electric field ${E_\perp = \beta_\mathrm{rot}B}$ and the parallel electric field $E_\parallel\sim E_\perp$ are typically much weaker than the magnetic field. Here $\beta_\mathrm{rot}\sim 10^{-6}$ is the speed at which a magnetic field line rotates due to the rotation of the magnetar. Thus the effects of the electric field are expected to contribute only small perturbations to the obtained results. The background electromagnetic 4-potential is chosen to be in the Landau gauge
\begin{equation}
    \mathcal{A}^\mu = (0,0,Bx,0).
\end{equation}

We use natural units, $\hbar=c=1$, where $\hbar$ is the reduced Planck constant and $c$ is the speed of light. In natural units, the fine-structure constant is $\alpha_e = \flatfrac{e^2}{(4\pi)}\simeq \flatfrac{1}{137}$, where $e$ is the elementary charge. The Dirac gamma matrices in the Dirac representation read
\begin{equation}
    \g{0}=\mqty(\mqty{1 & 0 \\ 0 & -1}),\quad \g{i} = \mqty(\mqty{0 & \sigma^i \\ -\sigma^i & 0}),
\end{equation}
where $\sigma^i$ are the Pauli matrices and each element represents a $2 \cross 2$ matrix. The Einstein summation convention is used, where repeated indices are summed over when one of them is upstairs and the other downstairs. Greek letters take the values $\{0,1,2,3\}$ or, equivalently, $\{t,x,y,z\}$ and are contracted with the Minkowski metric $g_{\alpha\beta}=\mathrm{diag}(1,-1,-1,-1)$, whereas Latin letters denote spatial indices and, thus, take the values $\{1,2,3\}$ and are contracted with the Euclidean metric $\delta_{ij}=\mathrm{diag}(1,1,1)$:
\begin{equation}
\begin{split}
    a^\alpha a_\alpha &= a\vdot a = (a^0)^2 - (a^1)^2 - (a^2)^2 - (a^3)^2 \\
    a^ia_i &= \Vec{a}\vdot\Vec{a} = (a^1)^2 + (a^2)^2 + (a^3)^2.
\end{split}
\end{equation}

Sometimes it is useful to differentiate between the parallel and perpendicular components of the 4-momentum, defined as: 
\begin{equation}
    \vec{k}_\parallel = \left(k^0,\; k^z\right),\;\vec{k}_\perp = \left(k^x,\; k^y\right).
\end{equation}
The parallel components mix together when a Lorentz transformation is applied in the $z$-direction, while the perpendicular components stay invariant. The scalar products of these vectors are defined as:
\begin{equation}
    k_\parallel^2 = \left(k^0\right)^2-\left(k^z\right)^2,\quad k_\perp^2=\left(k^x\right)^2+\left(k^y\right)^2.
\end{equation}
The Dirac-slash notation is used to denote contraction with the Dirac gamma matrices
\begin{equation}
    \slashed{A} = \g{\mu}A_\mu.
\end{equation}
Finally, the Fourier transform of $f(x)$ in 3+1-dimensional Minkowski spacetime is defined as 
\begin{equation}
    f(p)=\int \dd[4]{x} \e^{ip\vdot x} f(x)
\end{equation}
and the inverse Fourier transform as
\begin{equation}
    f(x)=\int \frac{\dd[4]{p}}{(2\pi)^4} \e^{-ip\vdot x} f(p).
\end{equation}

\subsection{\label{sec:FeynRules}Feynman rules}
The full Lagrangian of QED is
\begin{equation}
    \mathcal{L} = -\frac{1}{4}F_{\mu\nu}F^{\mu\nu} + \Bar{\psi}\left(\iu \slashed{D} - m_e\right)\psi,
\end{equation}
where $F_{\mu\nu}=\partial_\mu A_\nu - \partial_\nu A_\mu$ is the electromagnetic field tensor related to the dynamical gauge field $A_\mu$ and 
\begin{equation}
    D_\alpha = \partial_\alpha -\iu e \left(A_\alpha + \mathcal{A}_\alpha\right)
\end{equation}
is the gauge covariant derivative in the presence of a background electromagnetic field. Writing out the covariant derivative explicitly yields
\begin{equation}
    \mathcal{L} = -\frac{1}{4}F_{\mu\nu}F^{\mu\nu} + \Bar{\psi}\left(\iu\slashed{\partial} + e \slashed{\mathcal{A}} -m_e \right)\psi - q\Bar{\psi}\slashed{A}\psi.
\end{equation}
Thus, the Lagrangian splits into a sum of the quadratic Lagrangian of the free theory
\begin{equation}
    \mathcal{L}_0 = -\frac{1}{4}F_{\mu\nu}F^{\mu\nu} + \Bar{\psi}\left(\iu\slashed{\partial} + e \slashed{\mathcal{A}} -m_e \right)\psi
\end{equation}
and the interaction Lagrangian
\begin{equation}
    \mathcal{L}_\mathrm{int} = e\Bar{\psi}\slashed{A}\psi.
\end{equation}
We have chosen to include the interaction with $\mathcal{A}^\mu$ in the quadratic fermion part. The coupling of fermions to $\mathcal{A}^\mu$ is therefore treated non-perturbatively and exactly. This is also seen in our results that contain nonperturbative contributions from the magnetic field, e.g., $\exp(-k_\perp^2/(2b))$. 

\begin{figure}
    \centering
    \includegraphics[scale = 0.3]{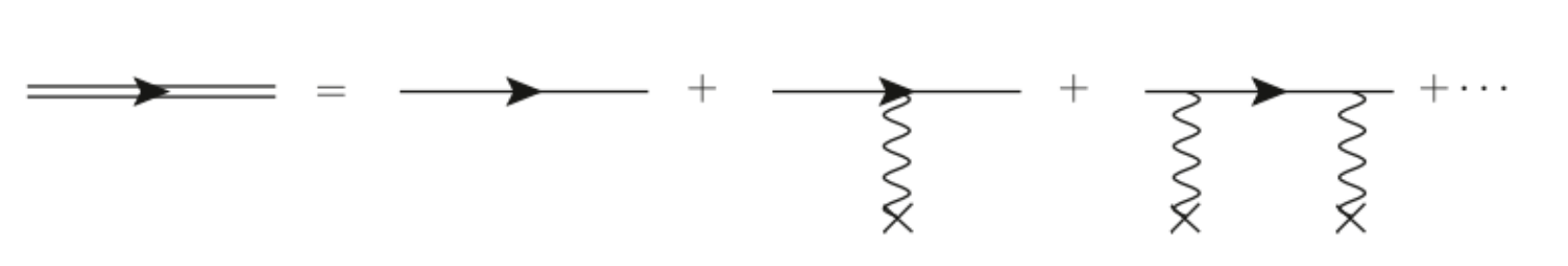}
    \caption{The fermion propagator in the Furry picture, denoted by a double line with an arrow in the middle, includes all possible interactions with the background magnetic field, denoted here by squiggly lines ending with a $\times$ \citep{fedotov_advances_2023}.}
    \label{fig:Furry}
\end{figure}

Compared to vacuum QED, the only difference is the addition of the background field in the quadratic fermion term. The effects of the background field are thus seen in the fermion propagator, while the photon propagator and interaction vertex work as in vacuum QED. The fermion propagator defined in this way can be thought of as including all possible interactions with the background field, as shown in Fig.~\ref{fig:Furry}. This way of presenting the fermion propagator is known as the Furry picture \cite{furry_bound_1951}. A strong background field introduces also other nonlinear QED effects, e.g., the Schwinger effect, in which electron--positron pairs are spontaneously created from the vacuum \cite{schwinger_gauge_1951}.

Recently, concerns have been raised about the region of validity of the Furry picture. For a review on the topic in the context of laser plasma physics, see, e.g., \cite{fedotov_advances_2023}. According to the Ritus-Narozhny conjecture \citep{narozhny_expansion_1980}, the effective expansion parameter of SFQED is
\begin{equation}
    \Tilde{\alpha} = \alpha \chi^{\frac{2}{3}},
\end{equation}
where
\begin{equation}
    \chi = \frac{e \sqrt{\abs{p^\mu F_{\mu\nu}}^2}}{m_e^3}
\end{equation}
is known as the quantum nonlinearity parameter and describes the strength of quantum effects on the particle. In the case of a constant background magnetic field, we obtain
\begin{equation}
    \chi_e = \sqrt{2nb^3}
\end{equation}
for an electron or positron. Thus, the analogous effective expansion parameter reads
\begin{equation}
    \tilde{\alpha} = \alpha b \sqrt[3]{2n}.
\end{equation}
Noteworthily, if a charged particle is constrained to the lowest Landau level, then $\chi_e=0$ and it is as if the effects of the magnetic field vanish. At the lowest Landau level the particle can only move along the magnetic field lines. This is an example of dimensional reduction, where a charged particle moving in a strong magnetic field in 3+1-dimensional spacetime, behaves like a free particle moving in 1+1-dimensional spacetime \citep{gusynin_dimensional_1995}. In contrast, for $n\gg 1$, also $\tilde{\alpha} \gg 1$ and the Furry picture breaks down as the perturbation expansion becomes invalid. In a strong magnetic field, electrons at high Landau levels de-excite rapidly through synchrotron radiation and thus the system self-regulates away from the $\tilde{\alpha} \gg 1$ region. For a photon the quantum non-linearity parameter reads
\begin{equation}
    \chi_\gamma = \frac{k_\perp b}{m_e}.
\end{equation}
For simplicity, I will not use the nonlinearity parameters in the subsequent derivations, but instead express the obtained cross sections in terms of $b$ and $k_\perp$.

Below, we list the Feynman rules of SFQED in a constant background magnetic field. Because $p^x$ is not well defined in the Landau gauge for electrons and positrons, the Feynman rules are given in position space. 

\subsubsection{External electron legs} \label{sec:electronFR}
An external electron leg in the Feynman diagram corresponds to the following expression in the scattering amplitude:
\begin{equation}
    \begin{split}
        \text{incoming electron: }& \us{n}{a}{\mu}{x}\e^{-\iu p\vdot x},\\
        \text{outgoing electron: }& \ubs{n}{a}{\mu}{x}\e^{\iu p\vdot x}=\us{n}{a}{\mu}{x}^\dagger\g{0}\e^{\iu p\vdot x}.
    \end{split}
\end{equation}
Here, $u$ is a spinor, $\mu = \pm 1$ labels the spin state of the particle, $n$ is the Landau level, $a=p^y/(qB)=s_\pm p^y\lb$ is the center of gyration, $s_\pm=\sign(qB)$, and $\lambda_B=\abs{qB}^{-\frac{1}{2}}$ is the magnetic length. The solution of the Dirac equation in a magnetic field is not a plane wave in the $x$-direction and therefore the $p$ in the exponent is $p^\mu=(E,0,p^y,p^z)$. 

We will use the transverse polarization states of the Sokolov--Ternov (ST) wave functions, first derived by \citet{sokolov_synchrotron_1968}. The ST wave functions are eigenfunctions of the $z$-component of the magnetic moment operator 

\begin{equation}
    \Vec{\mu}=m_e\Vec{\Sigma} - \iu\Vec{\gamma}\cross\Vec{p},\; \mu^z=\iu\g{1}\g{2}\left(m_e+\Vec{\gamma}_\perp\vdot \Vec{p}_\perp\right),
\end{equation}
where 
\begin{equation}
    \Sigma^i = \mqty(\mqty{\sigma^i & 0 \\ 0 & \sigma^i}).
\end{equation}
They were also derived by alternative reasoning by \citet{herold_cyclotron_1982}, who derived the ST wave functions by diagonalizing the interaction Hamiltonian that describes the coupling between the electrons and the background magnetic field. 

In early works on Compton scattering in a magnetic field (see, e.g., \citep{daugherty_compton_1986}) the Johnson--Lippmann (JL) wave functions were used. The JL wave functions are eigenfunctions of the kinetic momentum $\Vec{\Pi}$, however, they have a number of disadvantages compared to the ST wave functions. For example, they do not preserve the spin states of the electrons under Lorentz boosts along the magnetic field. For a comprehensive review on the differences between ST and JL wavefunctions see Ref.~\citep{gonthier_compton_2014}. A review of different wave function choices can be found in Ref.~\citep{melrose_quantum_1983}.

The general form of the electron wave function reads \citep{melrose_quantum_1983}
\begin{equation}
    \us{n}{a}{\mu}{x}=\mqty(\mqty{C_1\phi_{n-1,a}(x)  \\ C_2\phi_{n,a}(x) \\ C_3\phi_{n-1,a}(x) \\ C_4\phi_{n,a}(x)})\, ,
\end{equation}
where
\begin{equation}
    \phi_{n,a}(\Vec{x})=\frac{1}{L\left(\pi^\frac{1}{2}\lambda_B 2^n n!\right)^\frac{1}{2}}H_n\left(\frac{x-a}{\lambda_B}\right)\e^{-\frac{\left(x-a\right)^2}{2\lb}},
\end{equation}
and $H_n$ are the Hermite polynomials. To match with earlier work on SFQED in the context of astrophysics (e.g.~\citep{mushtukov_compton_2016, kostenko_qed_2018}), we have defined the wave function with periodic boundary conditions in a box with side length $L$. However, as will be shown in Sec.~\ref{sec:CrossSec}, no observables depend on $L$ and we will change our normalization such that $L=1$ starting from Sec.~\ref{sec:3part}.

The ST wave functions are given by \citep{kostenko_qed_2018}
\begin{equation}
\begin{split}
    \us{n}{a}{-1}{x} &= \frac{1}{f_n}\mqty(\mqty{-\iu p^z p_n \phi_{n-1,a}(x)  \\
    (E+E_{0})(E_{0}+m_e)\phi_{n,a}(x) \\
    -\iu p_n (E+E_{0})\phi_{n-1,a}(x) \\
    -p^z(E_{0}+m_e)\phi_{n,a}(x)})\, , \\
    \us{n}{a}{+1}{x} &= \frac{1}{f_n}\mqty(\mqty{(E+E_{0})(E_{0} + m_e)\phi_{n-1,a}(x)  \\
    -\iu p^z p_n\phi_{n,a}(x) \\
    p^z(E_{0}+m_e)\phi_{n-1,a}(x) \\
    \iu p_n (E+E_{0})\phi_{n,a}(x)})\, ,
\end{split}
\end{equation}
up to an arbitrary choice of phase. The wave functions use the shorthand notation 
\begin{equation}
\begin{split}
    p_n^2 &\equiv 2n\abs{qB} \equiv E_{0}^2-m_e^2, \\
    f_n &\equiv \sqrt{2E_{0}(E_{0}+m_e)(E_{0}+E)}.
\end{split}
\end{equation}
For $n=0$, all electrons are in the spin state $\mu=-1$,
\begin{equation}
    \us{0}{a}{-1}{x}=\mqty(\mqty{0 \\ C^\mathrm{ST}_2 \\ 0 \\ C^\mathrm{ST}_4})\phi_{0,a}(x)\equiv C^\mathrm{ST} \phi_{0,a}(x),
\end{equation}
where $C^\mathrm{ST}_2 = \sqrt{E+m_e}$ and $C^\mathrm{ST}_4=-\frac{p^z}{\sqrt{E+m_e}}$, and $H_0(x)=1$, which means that the electron wave function is a Gaussian,
\begin{equation}
    \phi_{0,a}(x)=L^{-1}\pi^{-\frac{1}{4}}\lambda_B^{-\frac{1}{2}}\e^{-\frac{\left(x-a\right)^2}{2\lb}}.
\end{equation}
The electron wave functions fulfill the following completeness and orthogonality conditions (${\Psi_{n,a}(\Vec{x}_\perp)=\phi_{n,a}(x)\e^{\iu p^y y},\; L=1}$):
\begin{equation} \label{eq:HermiteComp}
\begin{split}
    \int_{-\infty}^\infty \dd[2]{\Vec{x}_\perp}\Psi_{n,a}(\Vec{x}_\perp)\Psi_{m,a'}^*(\Vec{x}_\perp)&=2\pi\delta\left(p^y-(p')^y\right)\delta_{mn}, \\
    \int_{-\infty}^\infty \frac{\dd{p^y}}{2\pi}\sum_{n=0}^\infty \Psi_{n,a}(\Vec{x}_\perp)\Psi_{n,a}^*(\Vec{x}'_\perp)&=\delta^{(2)}\left(\Vec{x}_\perp-\Vec{x}_\perp'\right).
\end{split}
\end{equation}

\subsubsection{External positron legs}
External positron legs follow similar rules as electron legs:
\begin{equation}
    \begin{split}
        \text{incoming positron: }& \vbs{n}{a}{\mu}{x}\e^{-\iu p\vdot x},\\
        \text{outgoing positron: }& \vs{n}{a}{\mu}{x}\e^{\iu p\vdot x}.
    \end{split}
\end{equation}
The ST wave functions read \citep{kostenko_qed_2018}
\begin{equation}
\begin{split}
    v_{n,a}^{(+1)}(x) &= \frac{1}{f_n}\mqty(\mqty{- p_n (E+E_{0})\phi_{n-1,a}(x) \\
    -\iu p^z(E_{0}+m_e)\phi_{n,a}(x) \\
    -p^z p_n \phi_{n-1,a}(x) \\
    \iu (E+E_{0})(E_{0}+m_e)\phi_{n,a}(x)})\, , \\
    v_{n,a}^{(-1)}(x) &= \frac{1}{f_n}\mqty(\mqty{-\iu p^z(E_{0}+m_e)\phi_{n-1,a}(x) \\
    -p_n (E+E_{0})\phi_{n,a}(x) \\
    -\iu (E+E_{0})(E_{0} + m_e)\phi_{n-1,a}(x)  \\
    p^z p_n\phi_{n,a}(x)}).
\end{split}
\end{equation}
Positrons on the lowest Landau level are all in the spin state $\mu=+1$.

\subsubsection{External photon legs} \label{sec:photonFR}
An external photon leg corresponds to \citep{bjorken_relativistic_1965}:
\begin{equation}
\begin{split}
    \text{incoming photon: }& \varepsilon^\mu L^{-\frac{3}{2}}\e^{-\iu k\vdot x},\quad k^\nu = \omega(1, \hat{k}), \\
    \text{outgoing photon: }& \left(\varepsilon^\mu\right)^*L^{-\frac{3}{2}}\e^{\iu k\vdot x},
\end{split}
\end{equation}
where $\varepsilon^\mu$ is the polarization vector, $\omega$ is the angular frequency of the photon, and $\Vec{k}=\omega\hat{k}$ is the momentum of the photon. In the context of magnetized plasmas, the two photon-polarization modes used are the ordinary mode (O-mode) and the extraordinary mode (X-mode) \citep{arons_wave_1986}. The two modes are distinguished from each other by the direction of the electric field vector. In the O-mode, the electric field vector is parallel to $\Vec{k}\cross(\Vec{k}\cross\Vec{B})$, i.e.,~lies in the plane spanned by the vectors $\Vec{k}$ and $\Vec{B}$. On the other hand, in the X-mode, the electric field vector is parallel to $\Vec{k}\cross\Vec{B}$, i.e.,~it is perpendicular to the plane spanned by the vectors $\Vec{k}$ and $\Vec{B}$.

The polarization vectors of the two modes are \citep{kostenko_qed_2018}:
\begin{equation}
\begin{split}
    \varepsilon_\mathrm{O}^z&=\sin{\theta}, \quad\varepsilon_\mathrm{O}^\pm=\varepsilon_\mathrm{O}^x\pm\iu\varepsilon_\mathrm{O}^y=-\cos{\theta}\e^{\pm\iu\phi},\\
    \varepsilon_\mathrm{X}^z&=0,\quad\varepsilon_\mathrm{X}^\pm=\mp\iu\e^{\pm\iu\phi},
\end{split}
\end{equation}
where the angles give the propagation direction of the photon in spherical coordinates. These expressions are accurate up to corrections of order $\mathcal{O}(\sigma^{-1})$, where $\sigma = B^2/(4\pi n_\pm m_e)$ is the plasma magnetization, and $n_\pm$ is the number density of positrons and electrons, respectively. This means that the above polarization vectors are a good approximation for magnetic fields $b\lesssim 300$ \citep{mushtukov_compton_2016, medvedev_plasma_2023}. At larger magnetic fields, we would have to take into account changes in the vacuum dielectric tensor and the inverse permeability tensor caused by the magnetic field \citep{heyl_high-energy_2003,wang_photon_2021,medvedev_plasma_2023}.

For later convenience, it is also useful to define the functions
\begin{equation}
\begin{split}
    f_\mathrm{O}^z(\theta) &= \sin{\theta}, \quad f_\mathrm{O}^\pm(\theta) = -\cos{\theta}, \\
    f_\mathrm{X}^z(\theta) &= 0, \quad f_\mathrm{X}^\pm(\theta) = \mp\iu,
\end{split}
\end{equation}
i.e.,~the polarization vectors with the exponential function removed.

\subsubsection{Fermion--photon vertex}
The fermion--photon vertex is unchanged from vacuum QED and reads
\begin{equation}
    \text{Fermion-photon vertex: } -\iu q\g{\mu}\int\dd[4]{x}.
\end{equation}

\subsubsection{Fermion propagator}
The full derivation of the Schwinger-proper-time fermion propagator and its Landau level projection is given in Appendix \ref{app:ElecProp}. The Landau level projection of the propagator, given in a mixed representation, where a Fourier transform has been done for the $t$ and $z$ variables, reads
\begin{equation}
\begin{split} \label{eq:PropLevels}
    &S_F(E,p^z;\Vec{x}'_\perp, \Vec{x}_\perp)\equiv\iu \frac{\e^{\iu\Phi}\e^{-\frac{\xi^2}{2}}}{2\pi\lb} \sum_{n=0}^\infty \frac{F_n}{E^2-(p^z)^2-m_e^2-2n\abs{qB}}, \\
    &F_n=\left(\g{0}E-\g{3}p^z+m_e\right)\left( \mathcal{P}_{-}L_{n-1}(\xi^2) + \mathcal{P}_{+}L_n(\xi^2) \right) \\
    &+ \frac{\iu}{\lb} \Vec{\gamma}_\perp\vdot(\Vec{x}_\perp-\Vec{x}'_\perp)L_{n-1}^1(\xi^2).
\end{split}
\end{equation}
Here, $\Phi(\Vec{x}'_\perp, \Vec{x}_\perp)=-s_\pm\frac{(x+x')(y-y')}{2\lb}$ is the Schwinger phase, ${\xi^2=\frac{(x-x')^2 + (y-y')^2}{2\lb}}$, $\mathcal{P}_\pm~=~\frac{1}{2}\left(1\pm \iu s_\pm \g{1}\g{2} \right)$ are the spin projection operators, and $L_n(x)$ and $L_n^\alpha(x)$ are the Laguerre and generalized Laguerre polynomials, respectively. By convention, $L_{-1}^{\alpha}=0$. Note also that the order of $\vec{x}_\perp$ and $\vec{x}_\perp'$ is reversed compared to Eq.~\eqref{eq:PropLevelsApp}. The order chosen here represents a fermion propagating from a vertex at $x$ to a vertex at $x'$.

The sum over Landau levels in the above propagator can be eliminated to obtain the propagator in the Schwinger proper time method \citep{heisenberg_folgerungen_1936, schwinger_gauge_1951}. The calculation is done in App.~\ref{app:ElecProp} and the result reads
\begin{equation}
    \begin{split}
        S_F(E,p^z&;\Vec{x}'_\perp, \Vec{x}_\perp)=\frac{\e^{\iu\Phi}}{2\pi\lb} \int_0^\infty \dd{s} \e^{\iu s \left(E^2-(p^z)^2-m_e^2\right)}\e^{\iu\frac{\xi^2}{2}\cot(s\abs{eB})} \\
        &\cross\Big[ \left(\g{0}E-\g{3}p^z+m_e\right)\frac{\iu}{2}\left( s_\pm\g{1}\g{2} - \cot(s\abs{eB}) \right) \\
        &\quad\quad- \frac{\iu}{4\lb} \Vec{\gamma}_\perp\vdot(\Vec{x}_\perp-\Vec{x}'_\perp) \left( 1+\cot[2](s\abs{eB}) \right) \Big].
    \end{split}
\end{equation}
This version of the propagator has the advantage of not having an explicit sum over Landau levels. However, when we calculate cross sections we also need to include Landau level dependent decay rates into the propagator to regulate the resonances in the cross section (see Sec.~\ref{sec:damping} for further discussion on the topic). Thus, even though decay rates for the proper time propagator are available \citep{ayala_fermion_2021}, they do not correctly describe the damping of the resonances and it is vital to keep the Landau levels explicit in the propagator when calculating cross sections that include resonances \citep{ghosh_fermion_2024}. 

In previous studies on Compton scattering in astrophysical environments (see, e.g., \citep{mushtukov_compton_2016,gonthier_compton_2014,kostenko_qed_2018,melrose_quantum_1983-1,daugherty_compton_1986}) the following propagator has been used
\begin{equation}
\begin{split}
    S_F(x',x)&=-\iu\sum_{n_I,\mu_I}\int\frac{L\dd{a_I}}{2\pi\lb}\int\frac{L\dd{p^z_I}}{2\pi}\\
    &\Big[\theta\left(t'-t\right)\us{n_I}{a_I}{\mu_I}{\Vec{x}'}\ubs{n_I}{a_I}{\mu_I}{\Vec{x}}\e^{-\iu p_I\vdot \left(x'-x\right)} \\
    & - \theta\left(t-t'\right)\vs{n_I}{a_I}{\mu_I}{\Vec{x}'}\vbs{n_I}{a_I}{\mu_I}{\Vec{x}}\e^{\iu p_I\vdot \left(x'-x\right)} \Big],
    \end{split}
\end{equation}
where $\theta$ is the Heaviside step function. The choice of the propagator does not affect the obtained results, however, the right choice can make the subsequent calculations much less cumbersome. In this article, I will demonstrate how the Landau level projected Schwinger proper time propagator can be used to perform SFQED calculations.

\subsubsection{Photon propagator}
For the sake of completeness, the momentum space photon propagator in the Feynman gauge reads
\begin{equation}
    D_{\mu\nu}(p) = \frac{-\iu g_{\mu\nu}}{p^2}.
\end{equation}
I will not use the photon propagator in this article, but it is relevant for M\o ller and Bhabha scattering, which will be discussed in a follow-up paper. The Feynman rules of QED in a strong magnetic field are summarized in Table \ref{tab:FeynRules}.

\renewcommand{\arraystretch}{1.2}
\begin{table}[t]
    \centering
    \caption{Feynman rules for QED in a magnetic field.}\label{tab:FeynRules}
    \begin{tabular}{lc}
%    \hline\hline
       Part of diagram & Mathematical expression \\
       \hline
%       \hline
       Incoming electron  & $\us{n}{a}{\mu}{\Vec{x}}\e^{-\iu p\vdot x}$ \\
%       \hline
       Outgoing electron  & $\ubs{n}{a}{\mu}{\Vec{x}}\e^{\iu p\vdot x}$ \\
%       \hline %\vspace{0.4em}
       Incoming positron  & $\vbs{n}{a}{\mu}{\Vec{x}}\e^{-\iu p\vdot x}$ \\
%       \hline
       Outgoing positron  & $\vs{n}{a}{\mu}{\Vec{x}}\e^{\iu p\vdot x}$ \\
%       \hline
       Incoming photon & $\varepsilon^\mu\e^{-\iu k\vdot x}$ \\
%       \hline
       Outgoing photon & $(\varepsilon^\mu)^*\e^{\iu k\vdot x}$ \\
%       \hline
       Electron propagator & $S_F(E,p^z;\Vec{x}'_\perp, \Vec{x}_\perp)$ \\
%       \hline
       Photon propagator & $D_{\mu\nu}(p)$ \\
%       \hline
       Vertex & $-\iu q \g{\mu}\int\dd[4]{x}$ \\
%       \hline\hline
    \end{tabular}
\end{table}
\renewcommand{\arraystretch}{1}

\subsection{\label{sec:CrossSec}Calculating cross sections in a magnetic field}
Calculating cross sections in quantum field theories is a standard procedure covered in many textbooks \citep{peskin_introduction_1995, bjorken_relativistic_1965}. However, the presence of a magnetic field introduces some changes to the calculation. A review of these changes is given in Ref.~\citep{melrose_quantum_1983-1} and, in this section, I will give a brief overview on the topic.

All wave functions are defined with periodic boundary conditions in a cube with side length $L$ and temporal extent $T$. The transition rate $w_\mathrm{fi}$ from initial to final state is defined as
\begin{equation}
    w_\mathrm{fi} = \frac{\abs{T_\mathrm{fi}}^2}{L^3T}=(2\pi)^3\delta^{(3)}(E,p^z,p^y)\frac{\abs{M_\mathrm{fi}}^2}{L},
\end{equation}
where $T_\mathrm{fi} = (2\pi)^3\delta^{(3)}(E,p^z,p^y) M_\mathrm{fi}$ is the transition matrix of the interaction, and the relation
\begin{equation}
    \left[(2\pi)^3\delta^{(3)}(E,p^z,p^y)\right]^2 = L^2 T (2\pi)^3\delta^{(3)}(E,p^z,p^y)
\end{equation}
has been used. We have defined the shorthand notation
\begin{equation}
    \delta^{(3)}\left(E,p^z,p^y\right)\equiv \delta(E_\mathrm{in}-E_\mathrm{out})\delta(p^z_\mathrm{in}-p^z_\mathrm{out})\delta(p^y_\mathrm{in}-p^y_\mathrm{out})
\end{equation}
Note, that there is no conservation of momentum in the $x$-direction due to the electrons and positrons not having a well-defined $x$-momentum.

The cross section $\sigma$ of a scattering process reads
\begin{equation}
    \sigma = \int \prod_j \dd\left[\mathrm{PS}\right]_j \frac{w_\mathrm{fi}}{F},
\end{equation}
where $j$ runs over all outgoing particles, $\dd\left[\mathrm{PS}\right]_j$ is the density of states of outgoing particle $j$, and $F$ is the incoming particle flux. For photons the density of states reads
\begin{equation}
    \dd\left[\mathrm{PS}\right]_\gamma = L^3\frac{\dd[3]{k}}{2\omega(2\pi)^3} = L^3\frac{\omega\dd{\omega}\dd{\Omega}}{2(2\pi)^3},
\end{equation}
where the second expression is given in spherical coordinates. In the Landau gauge, electrons and positrons do not have a well-defined canonical momentum in the $x$-direction and, therefore, their density of states reads
\begin{equation}\label{eq:fermionPS}
    \dd\left[\mathrm{PS}\right]_{e^\pm} = L^2\frac{\dd[2]{p}}{2E(2\pi)^2} = L^2\frac{\dd{p^z}\dd{p^y}}{2E(2\pi)^2}.
\end{equation}
We choose to normalize our wave functions such that there are $2E$ particles per unit volume \citep{thomson_modern_2013, cannoni_lorentz_2017}. Then, the initial-state flux reads 
\begin{equation}
    F = 2E_\mathrm{in}L^{-3}
\end{equation}
for processes with one incoming particle and 
\begin{equation}
    F = 2E_\mathrm{in} 2E'_\mathrm{in}\Bar{v}L^{-6}
\end{equation}
for processes with two particles in the initial state. Here,
\begin{equation}\label{eq:MollerFlux}
    \Bar{v} = \sqrt{\left(\Vec{v}-\Vec{v}'\right)^2-\left(\Vec{v}\cross\Vec{v}'\right)^2}
\end{equation}
is the relativistic generalization of the relative velocity of the two incoming particles (sometimes referred to as the M\o ller velocity) \cite{cannoni_lorentz_2017} and $\Vec{v}$ and $\Vec{v}'$ are the velocities of the two incoming particles. Additionally, since the $p^y$ of an incoming electron or positron is not known, we must average over it by integrating the cross section over all possible values of $p^y$,
\begin{equation}
    \bar{\sigma}=\frac{\lb}{L}\int\dd{p^y}\sigma.
\end{equation}
The factor of $\lb L^{-1}$ in front of the integral is a normalization factor that is obtained from the density of states, i.e.,
\begin{equation}
    \int \dd{p^y} = \frac{L}{\lb}.
\end{equation}
It is also important to note that $p^y$ is a gauge dependent quantity. The averaging process ensures that our results are gauge invariant. 

\begin{figure}
    \includegraphics[width = .7\textwidth]{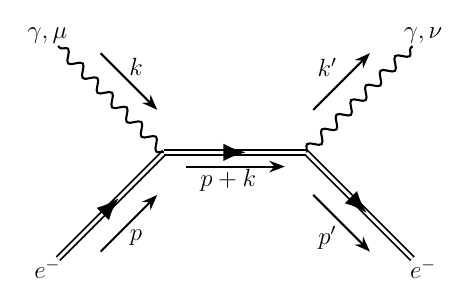}
    \caption{Feynman diagram of Compton scattering. Electrons are denoted by $e^-$ and are drawn as double lines with an arrow in the middle. Photons are denoted by $\gamma$ and are drawn as squiggly lines. The Lorentz index of their polarization vectors is written next to the $\gamma$. The direction of the momentum of the particles is denoted by external arrows. }
    \label{fig:FeynEx}
\end{figure}

For example, the cross section of Compton scattering (Feynman diagram shown in Fig.~\ref{fig:FeynEx}) is given by
\begin{equation}
\begin{split}
    \bar{\sigma} &= \frac{\lb}{L}\int \dd{p^y} \int L^2\frac{\dd[2]{p'}}{(2\pi)^2}L^3\frac{\dd[3]{k'}}{(2\pi)^3}\frac{w_\mathrm{fi}}{F} \\
    &= \frac{\lb L^{9}}{\bar{v}(2\pi)^2}\int\dd{p^y}\dd[2]{p'}\dd[3]{k'}\delta^{(3)}(E,p^z,p^y)\abs{M_\mathrm{fi}}^2.
\end{split}
\end{equation}
This integral is evaluated in Sec. \ref{sec:Compton}. Importantly, the integrand is independent of $p^y$ and, thus, the integration over $p^y$ yields a multiplicative factor of $L \lambda_B^{-2}$. Therefore, since the squared matrix element is proportional to $\abs{M_\mathrm{fi}}^2 \propto L^{-10}$, the cross section is independent of the normalization length $L$ and time $T$, as expected. Similarly, all other cross sections are also independent of the normalization scales. For the sake of brevity, we will adopt the convention $L=T=1$ in our calculations.

Finally, a few remarks on the Lorentz invariance of the theory. Electromagnetic fields are invariant under Lorentz boosts in the direction of the field \citep{griffiths_introduction_2017}. The only electromagnetic field present in our calculations is a constant magnetic field in the $z$-direction. Thus, we are limited to Lorentz boosts in the $z$-direction to keep the field invariant.

It is a well-known fact that the integration measure $\frac{\dd[3]{p}}{2E}$ is Lorentz invariant \citep{thomson_modern_2013}. However, in our calculations we instead use the integration measure $\frac{\dd[2]{p}}{2E}$, as given in Eq.\eqref{eq:fermionPS}. Once again Lorentz invariance is saved if we restrict ourselves to boosts in the $z$-direction because such transformations only affect the temporal and $z$-components of the measure, which are the same as in the vacuum case.

With the chosen normalization for the wavefunctions both the matrix element $M_\mathrm{fi}$ and incoming particle flux $F$ are also Lorentz invariant under boosts in the $z$-direction \citep{thomson_modern_2013}. Therefore,  all components of the cross section are invariant and thus also the cross section is invariant under Lorentz boosts in the $z$-direction and we are free to choose the frame that is most beneficial for our calculations. We will mainly use two different Lorentz frames---the electron rest frame (ERF), where the incoming electron has no $z$-momentum, and the center-of-$z$-momentum (Co$z$M) frame, where the total $z$-momentum of the system is zero.

\section{\label{sec:3part}$\mathcal{O}(\alpha_e)$ cross sections}

At tree level, all 3-particle QED scattering processes consist of a single vertex between a photon and two charged fermions. Many of the cross sections of 3-particle scattering processes have already been calculated in the literature. However, the results are often given for spin and polarization averaged cross sections and are scattered across many different papers. Here I demonstrate that our formalism can reproduce the known results and provide simple access to cross sections of specific spin and polarization states.

No explicit expressions are given for the cross sections due to the amount of different cases and the length of the expressions. Instead, I present the necessary steps required to perform the calculation. The cross sections are made available for everyone via a Python package available on GitHub~\footnote{\url{https://github.com/hel-astro-lab/SFQED}\label{fn:github}}. We will focus mainly on synchrotron radiation---the scattering amplitudes of all other processes can be obtained from the results of synchrotron radiation via crossing symmetry.

\subsection{\label{sec:Synchrotron}Synchrotron radiation}
\begin{figure}
    \centering
    \includegraphics[width = .6\textwidth]{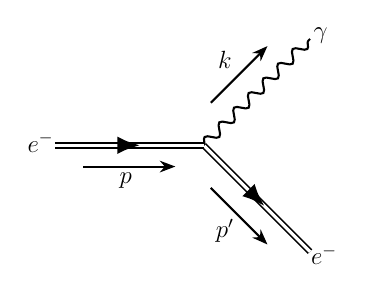}
    \caption{Feynman diagram of synchrotron radiation.}
    \label{fig:Synchrotron}
\end{figure}
The tree-level Feynman diagram of synchrotron radiation is shown in Fig. \ref{fig:Synchrotron}. Using the Feynman rules given in Sec. \ref{sec:FeynRules}, we obtain the scattering amplitude for synchrotron radiation
\begin{equation}
\begin{split}
    \iu T_\mathrm{SR} &= -\iu e\int\dd[4]{x}\ubs{n'}{a'}{\sigma'}{x} \slashed{\varepsilon}_k^*\us{n}{a}{\sigma}{x} \e^{\iu\left(p'+k-p\right)\vdot x} \\
    &= -\iu e (2\pi)^3\delta^{(3)}(E,p^z,p^y) \int\dd{x} \ubs{n'}{a'}{\sigma'}{x} \slashed{\varepsilon}_k^*\us{n}{a}{\sigma}{x} \e^{-\iu k^x x}. \\
\end{split}
\end{equation}
The evaluation of the final integral is done in App. \ref{app:SynchrotronApp}.

The decay rate of synchrotron radiation is most convenient to calculate in the rest frame of the electron. Following the rules given in Sec.~\ref{sec:CrossSec}, the synchrotron decay rate reads
\begin{equation}
    \begin{split}
        \Gamma_\mathrm{SR} &= \lb\int \dd{p^y} \int \frac{\dd[2]{p'}}{2E'(2\pi)^2}\frac{\dd[3]{k}}{2\omega}\delta^{(3)}(E,p^z,p^y)\frac{\abs{M_\mathrm{SR}}^2}{2E} \\
        &= \frac{1}{8\pi} \int\dd(\cos{\theta}) \frac{\omega}{E' + \omega\cos[2](\theta)}\frac{\abs{M_\mathrm{SR}}^2}{2E},
    \end{split}
\end{equation}
where we have made use of the fact that the squared matrix element does not depend on the azimuthal angle $\phi$ due to cylindrical symmetry. The decay rates at nonzero $p^z$ are related to the stationary case by a Lorentz transformation 
\begin{equation}\label{eq:DecayBoost}
    \Gamma(p^z) = \frac{\sqrt{m^2+2eBn}}{\sqrt{(p^z)^2+m^2+2eBn}}\Gamma(p^z=0).
\end{equation}
The resulting cross section matches the one obtained in Ref.~\cite{herold_cyclotron_1982} after averaging over outgoing spin and polarization states.

\subsection{One-photon pair creation}
The squared matrix element for one-photon pair creation is obtained by the same calculation as for synchrotron radiation, but with the substitutions:
\begin{equation}
    \begin{split}
        &p^\mu \to -p^\mu,\quad k^\mu\to -k^\mu, \\
        &\us{n}{a}{\sigma}{x} \to \vs{n}{a}{\sigma}{x},\quad \varepsilon_k^* \to \varepsilon_k.
    \end{split}
\end{equation}
Thus, the decay rate of one-photon pair creation in a general reference frame reads
\begin{equation}
    \begin{split}
        \Gamma_\mathrm{1PC} &= \int \frac{\dd[2]{p}}{2E(2\pi)}\frac{\dd[2]{p'}}{2E'}\delta^{(3)}(E,p^z,p^y)\frac{\abs{M_{1PC}}^2}{2\omega} \\
        &= \sum_{p^z} \frac{1}{2\pi\lb} \abs{\frac{p^z}{E} - \frac{(p')^z}{E'}}^{-1}\frac{\abs{M_{1PC}}^2}{8\omega E E'}.
    \end{split}
\end{equation}
Importantly, there are two values of $p^z$ that satisfy the conservation laws and we must therefore sum over them. Note also that
\begin{equation}
    \abs{\frac{p^z}{E} - \frac{(p')^z}{E'}} = \abs{v_z-v'_z}
\end{equation}
is the difference of the velocities of the produced electron and positron along the magnetic field line. Compared to synchrotron radiation, the decay rate of one-photon pair creation has an additional factor of $\lambda_B^{-2}\sim b$. Thus, the decay rate is amplified in a strong magnetic field. The obtained cross section matches with the previously known result when the fermions are restricted to the LLL \citep{kostenko_qed_2018}.

\subsection{One-photon pair annihilation}
The squared matrix element for one-photon pair annihilation is obtained from the synchrotron radiation calculation with the substitutions
\begin{equation}
    (p')^\mu \to -(p')^\mu,\quad \ubs{n'}{a'}{\sigma'}{x} \to \vbs{n'}{a'}{\sigma'}{x}.
\end{equation}
The cross section is most convenient to calculate in the Co$z$M frame, but can also be calculated in, e.g., the ERF. The electron and positron are both moving along the magnetic field line (we ignore the gyration around the magnetic field line when calculating the relative speed of the particles). The cross section reads
\begin{equation}
    \begin{split}
        \sigma_\mathrm{1PA} &= \frac{\lambda_B^4}{\Bar{v}}\int\dd{p^y}\dd{(p')^y} \int \frac{\dd[3]{k}}{2\omega}\delta^{(3)}(E,p^z,p^y)\frac{\abs{M_{1PA}}^2}{4EE'} \\
        &= \frac{2\pi\lb}{\Bar{v}}\frac{\abs{M_{1PA}}^2}{8EE'},
    \end{split}
\end{equation}
where we have made use of the fact that the squared matrix element does not depend on the azimuthal angle $\phi$ due to rotational symmetry. The cross section has an additional factor of $\lb\sim b^{-1}$ compared to synchrotron radiation and is thus suppressed for strong magnetic fields. For both fermions on the LLL our results match with the known cross section\footnote{There is a missing factor of $4\pi^2$ in eq.~(11) of Ref.~\citep{daugherty_pair_1980}.} \citep{daugherty_pair_1980, wunner_comparison_1979}.

\subsection{Synchrotron self-absorption}
The squared matrix element for synchrotron self-absorption (SSA) is obtained by the same calculation as for synchrotron radiation, but with the substitutions
\begin{equation}
    k^\mu \to -k^\mu,\quad \varepsilon_k^* \to \varepsilon_k.
\end{equation}
The cross section is most conveniently calculated in the rest frame of the incoming electron, in which case $\Bar{v}=1$.
\begin{equation}
    \begin{split}
        \sigma_\mathrm{SSA} &= 2\pi\lb\int\dd{p^y} \int \frac{\dd[2]{p'}}{2E'}\delta^{(3)}(E,p^z,p^y)\frac{\abs{M_{SA}}^2}{4E\omega} \\
        &= 2\pi \delta(E+\omega-E')\frac{\abs{M_{SA}}^2}{8EE'\omega}.
    \end{split}
\end{equation}

Unlike the other $\mathcal{O}(\alpha_e)$ cross sections, the cross section of SSA contains a $\delta$-function that enforces the conservation of energy. In practice this means that SSA is only possible when
\begin{equation}
    n_\mathrm{SSA} \equiv \frac{1}{2b}\left(\left(\frac{k_\perp}{m_e}\right)^2+\frac{2\omega}{m_e}\sqrt{1+2n_\mathrm{i} b}\right) + n_\mathrm{i}
\end{equation}
is an integer. Due to the inclusion of the $\delta$-function, the cross section for SSA has not been implemented into our Python package.

\section{\label{sec:4part}$\mathcal{O}(\alpha_e^2)$ cross sections}

In this section, I compute the cross sections of all tree-level 2-to-2-particle scattering processes that have a fermion propagator in their Feynman diagram. Unlike in the previous section, the full scattering amplitudes of Compton scattering are given in App.~\ref{app:ComptonResult}. The complete analytical expressions of these scattering amplitudes have not been published before. The scattering amplitudes of the other scattering processes are made available through our Python package\footnotemark[1]. The focus in this section is on calculating the cross section of Compton scattering. The cross sections of the other processes can be obtained via crossing symmetry. 

\subsection{\label{sec:damping}Decay rates}
\begin{figure}
    \centering
    \includegraphics[width = .9\textwidth]{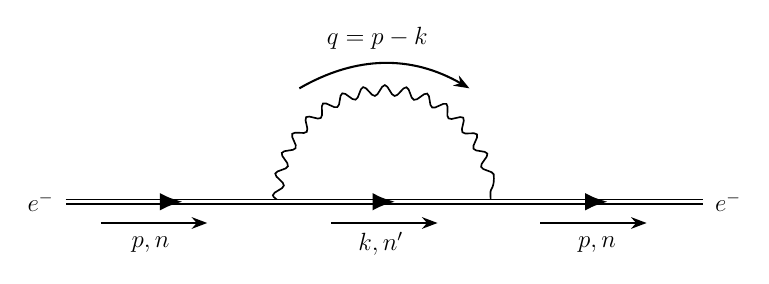}
    \caption{Fermion self-energy at 1-loop order.}
    \label{fig:SelfEnergy}
\end{figure}

The cross sections of 2-to-2-particle scattering processes contain resonances, i.e., infinite peaks, that are caused by the fermion propagator going on-shell. Traditionally in astrophysical contexts, the infinite resonances have been regulated by implementing an imaginary shift to the energy of the propagator
\begin{equation} \label{eq:EnergyShift}
    E \to E -\frac{\iu\Gamma}{2},
\end{equation}
where $\Gamma$ is the synchrotron decay rate calculated in Sec. \ref{sec:Synchrotron} summed over spin and polarization states of the outgoing electron and photon, respectively \cite{herold_cyclotron_1982, gonthier_compton_2014,mushtukov_compton_2016}. However, while this traditional approach is satisfactory in most use cases, it hides the true origin of the regulator. In QFT, the correct way to regulate divergences is by calculating higher-order perturbative corrections to the propagator, i.e., the imaginary part of the electron self-energy \cite{graziani_strong-field_1993,ayala_fermion_2021, ghosh_fermion_2024}. At 1-loop-order, the electron self-energy is given by the Feynman diagram shown in Fig. \ref{fig:SelfEnergy}. Unlike the synchrotron decay rate calculation, the self-energy calculation is straightforward to generalize to, e.g., non-zero temperature \cite{ghosh_fermion_2024}.

To correctly implement the different decay rates of the two spin states, the electron propagator must be split using the spin projection operators $\mathcal{P}_\pm$
\begin{equation}
\begin{split} 
    S_\mathrm{F}(E,p^z&;\Vec{x}'_\perp, \Vec{x}_\perp)=\iu \frac{\e^{\iu\Phi(x',x)}\e^{-\frac{\xi^2}{2}}}{2\pi\lb}\\
    &\cross\sum_{n=0}^\infty \Bigg(\frac{F_n^+}{E^2-(p^z)^2-\left(m_e-\frac{\iu\Gamma^+}{2}\right)^2-2n\abs{qB}} \\
    &+ \frac{F_n^-}{E^2-(p^z)^2-\left(m_e-\frac{\iu\Gamma^-}{2}\right)^2-2n\abs{qB}}\Bigg),
\end{split}
\end{equation}
where
\begin{equation}
\begin{split}
    F_n^+&=\mathcal{P}_+\Bigg(\left(\g{0}E-\g{3}p^z+m_e\right)L_n\left(\xi^2\right) \\
    &+ \frac{\iu}{\lb} \Vec{\gamma}_\perp\vdot(\Vec{x}_\perp-\Vec{x}'_\perp)L_{n-1}^1\left(\xi^2\right)\Bigg),\\
    F_n^-&=\mathcal{P}_-\Bigg(\left(\g{0}E-\g{3}p^z+m_e\right)L_{n-1}\left(\xi^2\right)\\
    &+ \frac{\iu}{\lb} \Vec{\gamma}_\perp\vdot(\Vec{x}_\perp-\Vec{x}'_\perp)L_{n-1}^1\left(\xi^2\right)\Bigg),
\end{split}
\end{equation}
and the completeness property $\mathcal{P}_+ + \mathcal{P}_- = 1$ has been used.

\begin{figure}
    \centering
    \includegraphics[scale = 0.6]{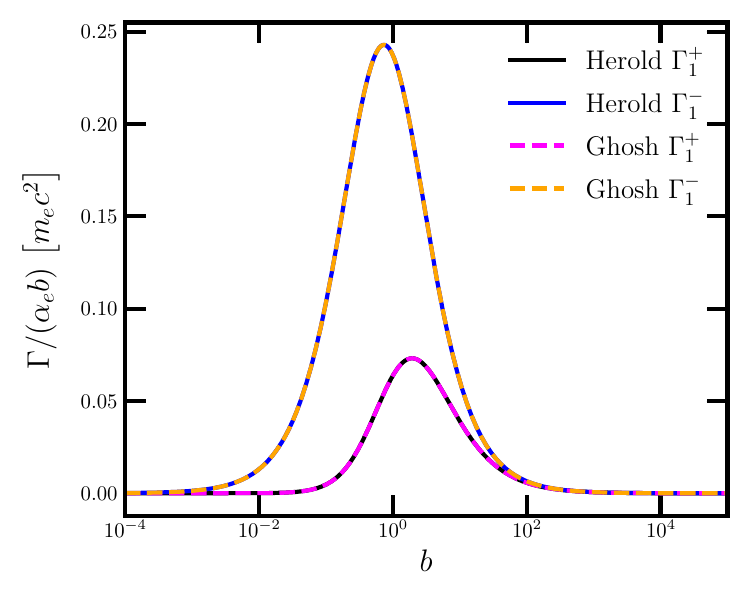}
    \caption{Spin-dependent decay rate of the first excited Landau level calculated from the electron self-energy (dashed lines, \citet{ghosh_fermion_2024}) and synchrotron radiation (solid lines, \citet{herold_cyclotron_1982}), respectively.}
    \label{fig:Damping}
\end{figure}

I have calculated the imaginary part of the electron self-energy by taking the zero-temperature limit of the results obtained in Ref.~\citep{ghosh_fermion_2024}. A brief overview is given in App.~\ref{app:DampingApp}. The obtained decay rate matches perfectly with the spin and polarization averaged synchrotron decay rate in the limit where the external electron legs of the electron self-energy diagram go on-shell, as shown in Fig.~\ref{fig:Damping}. The external electron legs going on-shell in the self-energy diagram corresponds to the propagator going on-shell and since the main contribution of the decay rate to the cross sections is near the resonances, the use of the synchrotron decay rate is a reasonable approximation. However, it should be noted that the effect of the fermion self-energy on the propagator is to shift the position of the poles of the propagator \citep{ghosh_fermion_2024}. Thus, the decay rate should be implemented as a shift in the electron mass in the denominator of the propagator, $m\to m -\iu\Gamma/2$, instead of the energy shift given in Eq.~\eqref{eq:EnergyShift}. We have implemented both ways of calculating the decay rate into our Python package. 

\subsection{\label{sec:Compton}Compton scattering}
\begin{figure}
\sidesubfloat[]{
\label{fig:SDiag}\includegraphics[width = 0.5\textwidth, trim = {0 0 0 -5ex},clip]{Figures/compton_labels.pdf}}
\hfil
\sidesubfloat[]{\label{fig:UDiag}\includegraphics[width = .3\textwidth]{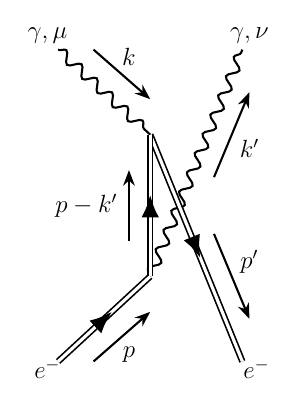}}
\hfil
    \caption{Tree level Compton scattering Feynman diagrams for the (a) $s$-channel and (b) $u$-channel.}
    \label{fig:CompDiag}
\end{figure}

The cross section of Compton scattering in a strong background magnetic field, taking into account all Landau levels of both the external electrons and the propagator was first calculated in Ref.~\cite{mushtukov_compton_2016}. In this section, I present the same calculation with the new formalism. The results agree reasonably well with the known cross section aside from a difference in how the decay widths are implemented for external fermions. See the companion letter \citep{kiuru_qed_2026} for a more detailed comparison.

The Compton scattering cross section is calculated in the rest frame of the incoming electron, where $\Bar{v} = 1$. The cross section reads
\begin{equation}
    \begin{split}
        \sigma_\mathrm{C} &= \lb\int\dd{p^y} \int \frac{\dd[2]{p'}}{2E'(2\pi)^2}\frac{\dd[3]{k'}}{2\omega'}\delta^{(3)}(E,p^z,p^y)\frac{\abs{M_\mathrm{C}}^2}{4E\omega} \\
        &= \frac{1}{(2\pi)^2} \int\dd{\Omega}\frac{\omega'}{E' - (p')^z\cos{\theta'}}\frac{\abs{M_\mathrm{C}}^2}{16E\omega}.
    \end{split}
\end{equation}

There are two Feynman diagrams contributing to the process---the $s$-channel and the $u$-channel shown in Fig.~\ref{fig:CompDiag}. However, we do not need to explicitly calculate both channels sinc the $u$-channel amplitude corresponds to the substitutions $k^\mu \leftrightarrow -(k')^\mu$ and $\varepsilon_k \leftrightarrow \varepsilon_{k'}^*$ in the $s$-channel amplitude. 

There are three different levels of accuracy when calculating the Compton scattering cross section. The first level is the lowest Landau level approximation (LLLa), where all fermions (external and virtual) are confined to the lowest Landau level. The calculation of the Compton scattering cross section in the LLLa is done in Appendix \ref{app:LLL} and the result matches with earlier calculations \citep{kostenko_qed_2018}. The LLLa is, however, only valid when the magnetic field is the dominating energy scale, i.e, $k_\perp \lesssim m_e\sqrt{b}$. For a more in-depth discussion on the topic, see the companion letter \citep{kiuru_qed_2026}.

The next level of approximation is to allow for the virtual fermions to occupy any Landau level, while still requiring that the external electrons stay on the lowest Landau level. Then, the scattering amplitudes read
\begin{widetext}
\begin{equation}
    \begin{split}
    \iu T_s &= (-\iu e)^2\int \dd[4]{x'}\int \dd[4]{x}\ubs{0}{a'}{-1}{x'}\e^{\iu p'\vdot x'}\slashed{\varepsilon}^*_{k'}\e^{\iu k'\vdot x'} S_\mathrm{F}(x',x)\slashed{\varepsilon}_k\e^{-\iu k\vdot x}\us{0}{a}{-1}{x}\e^{-\iu p\vdot x} \equiv \iu\sum_{n=0}^\infty T_s^{(n)} \\
    \end{split}
\end{equation}
and
\begin{equation}
    \begin{split}
    \iu T_u &= (-\iu e)^2\int \dd[4]{x'}\int \dd[4]{x}\ubs{0}{a'}{-1}{x'}\e^{\iu p'\vdot x'}\slashed{\varepsilon}_k\e^{-\iu k\vdot x'} S_\mathrm{F}(x',x)\slashed{\varepsilon}^*_{k'}\e^{\iu k'\vdot x}\us{0}{a}{-1}{x}\e^{-\iu p\vdot x} \equiv \iu\sum_{n=0}^\infty T_u^{(n)},\\
    \end{split}
\end{equation}
respectively. Thus, the evaluation of the scattering amplitudes boils down to calculating the integrals $T_s^{(n)}$ and $T_u^{(n)}$. We will focus on the $s$-channel and obtain the result of the $u$-channel via symmetry.

Let us calculate the contribution of the $n$th term to the sum in the $s$-channel. The calculation proceeds as in the LLLa, but replacing $S_{\mathrm{F},0}(E_I,p^z_I; \Vec{x}'_\perp, \Vec{x}_\perp)$ with $S_{\mathrm{F},n}(E_I,p^z_I; \Vec{x}'_\perp, \Vec{x}_\perp)$. This yields
\begin{equation}
\begin{split}
    \iu T_s^{(n)} &= -\frac{2\pi \iu e^2}{\pi^\frac{1}{2}\lambda_B^3}\delta(E + \omega - E' - \omega')\delta(p^z+k^z-(p')^z-(k')^z)\int \dd[2]{x_\perp}\dd[2]{x_\perp'}\e^{\iu (p^y+k^y)y}\e^{-\iu \left((p')^y+(k')^y\right)y'}\e^{\iu k^x x}\e^{-\iu (k')^x x'} \\
    &\cross \e^{-\frac{(x-a)^2}{2\lb}}\e^{-\frac{(x'-a')^2}{2\lb}}\e^{-\iu s_\pm\frac{(x+x')(y-y')}{2\lb}}\e^{-\frac{(x-x')^2+(y-y')^2}{4\lb}}\Big(\Bar{A}^+ A^+ L_n\left(\xi^2\right) + \Bar{A}^- A^- L_{n-1}\left(\xi^2\right)+ \frac{\iu}{\lb}\Bar{B}\left(\Vec{\gamma}_\perp \vdot (\Vec{x}_r)_\perp\right) B L_{n-1}^1\left(\xi^2\right)\Big)\\
    &\equiv I_{A^+}^{(n)} + I_{A^-}^{(n-1)} + I_B^{(n-1)},
\end{split}
\end{equation}
\end{widetext}
where
\begin{equation}
    \begin{split}
        &\Bar{A}^+ A^+ \equiv\left(C^{\mathrm{ST'}}\right)^\dagger\g{0}\slashed{\varepsilon}_{k'}^*\frac{\left(\g{0}E_I-\g{3}p_I^z+m_e\right)}{E_I^2-\left(p_I^z\right)^2-m_e^2-2eBn} \mathcal{P}_+\slashed{\varepsilon}_kC^{\mathrm{ST}},\\
        &\Bar{A}^- A^- \equiv \left(C^{\mathrm{ST'}}\right)^\dagger\g{0}\slashed{\varepsilon}_k\frac{\left(\g{0}E_I-\g{3}p_I^z+m_e\right)}{E_I^2-\left(p_I^z\right)^2-m_e^2-2eBn} \mathcal{P}_-\slashed{\varepsilon}_{k'}^*C^{\mathrm{ST}},\\
        &\Bar{B}\left(\Vec{\gamma}_\perp \vdot \Vec{x}_r\right) B \equiv \left(C^{\mathrm{ST'}}\right)^\dagger\g{0}\slashed{\varepsilon}_{k'}^*\frac{\left(\Vec{\gamma}_\perp \vdot \Vec{x}_r\right)}{E_I^2-\left(p_I^z\right)^2-m_e^2-2eBn} \slashed{\varepsilon}_kC^{\mathrm{ST}}.
    \end{split}
\end{equation}

Note that the (generalized) Laguerre polynomials are independent of $X$. Thus, it is reasonable to perform the integral over $X$ first. The first integral yields
\begin{equation}
    \begin{split}
        &I_{A^+}^{(n)}=-\frac{(2\pi)^2 \iu e^2}{\lb}\delta^{(3)}(E,p^z,p^y)\Bar{A}^+ A^+ \int \dd{x_r}\dd{y_r}\e^{-\frac{x_r^2+y_r^2}{2\lb}} \\
        &\cross \e^{-\frac{\lb}{4} \left(2\left(k^y-(k')^y\right)^2 + \left( k^x -(k')^x +\iu \left( k^y - (k')^y \right) \right)^2\right)} \e^{\iu y_r \upsilon_y} \e^{\iu x_r \upsilon_x} L_n\left(\xi^2\right),
    \end{split}
\end{equation}
where
\begin{equation}
\begin{split}
    \upsilon_x &\equiv \frac{1}{2}\left( k^x+(k')^x -\iu \left(k^y-(k')^y\right)\right), \\
    \upsilon_y &\equiv \frac{\iu}{2}\left( k^x -(k')^x -\iu \left( k^y + (k')^y \right) \right),
\end{split}
\end{equation}
and we have used the remaining gauge freedom to set $p^y=0$ to simplify the calculation.

After the change of variables into polar coordinates,
\begin{equation}
\begin{split}
    r^2&=x_r^2+y_r^2, \quad \Vec{x}_r\vdot\Vec{\upsilon}=r\upsilon\cos{\varpi}, \\
    \upsilon^2 &= \upsilon_x^2 + \upsilon_y^2 = \omega\omega'\sin{\theta}\sin{\theta'}\e^{\iu(\phi'-\phi)},
\end{split}
\end{equation}
the integral $I_{A^+}$ simplifies to (dropping the multiplicative factor in front of the integral for notational simplicity)
\begin{equation}
    \begin{split}
        I_{A^+}^{(n)}&\sim\e^{-\frac{\lb}{4} \left(2\left(k^y-(k')^y\right)^2 + \left( k^x -(k')^x +\iu \left( k^y - (k')^y \right) \right)^2\right)} \\
        &\cross \int_0^\infty \dd{r} r \e^{-\frac{r^2}{2\lb}} L_n\left(\frac{r^2}{2\lb}\right) \int_0^{2\pi} \dd{\varpi} \e^{\iu r\upsilon \cos{\varpi}} \\
        &= 2\pi\lb\e^{-\frac{1}{4}\lb\left(k_\perp^2 + (k')_\perp^2\right)} \e^{\iu \Theta} \e^{\iu \alpha} \frac{\left(\lb k_\perp k_\perp'\e^{\iu(\phi'-\phi)}\right)^n}{2^n n!}.
    \end{split}
\end{equation}
Here we have used the known integral \citep[eq. 7.421.1]{gradshtein_table_2015},
\begin{equation} \label{eq:LaguerreInt}
\begin{split}
    &\int_0^\infty \dd{x} x \e^{-\frac{1}{2}\vartheta x^2} L_n\left(\frac{1}{2}\beta x^2\right) J_0\left(xy\right) \\
    &= \frac{\left(\vartheta-\beta\right)^n}{\vartheta^{n+1}}\e^{-\frac{1}{2\vartheta}y^2} L_n\left(\frac{\beta y^2}{2\vartheta (\beta-\vartheta)}\right),
\end{split}
\end{equation}
with $\vartheta=\beta=1$, recalling that the leading order behavior of the Laguerre polynomials is $L_n(x)\sim (-1)^n\flatfrac{x^n}{n!}\binoppenalty=10000 
\relpenalty=10000$. The variables $\Theta$ and $\alpha$ are defined in Eq.~\eqref{eq:ThetaAlpha}.
The final expression reads
\begin{equation}
    \begin{split}
        I_{A^+}^{(n)} &= -(2\pi)^3\iu e^2\delta^{(3)}(E,p^z,p^y)\e^{-\frac{1}{4}\lb\left(k_\perp^2 + (k')_\perp^2\right)} \e^{\iu \Theta} \e^{\iu \alpha} \\
        &\cross \Bar{A}^+ A^+ \frac{\left(\lb k_\perp k_\perp'\e^{\iu(\phi'-\phi)}\right)^n}{2^n n!}.
    \end{split}
\end{equation}
The calculation of $I_{A^-}^{(n-1)}$ can be done in exactly the same way.

A similar result can be obtained for
\begin{equation}
    \begin{split}
        &I_B^{(n)}=\frac{(2\pi)^2e^2}{\lambda_B^4}\delta^{(3)}(E,p^z,p^y) \int \dd{x_r}\dd{y_r}\e^{-\frac{x_r^2+y_r^2}{2\lb}}\e^{\iu y_r \upsilon_y} \e^{\iu x_r \upsilon_x} \\
        &\cross\e^{-\frac{\lb}{4} \left(2\left(k^y-(k')^y\right)^2 + \left( k^x -(k')^x +\iu \left( k^y - (k')^y \right) \right)^2\right)}  \Bar{B}\left(\Vec{\gamma}_\perp \vdot (\Vec{x}_r)_\perp\right) B L_n^1\left(\xi^2\right).
    \end{split}
\end{equation}
Changing to polar coordinates (again dropping the factor in front of the integral) yields
\begin{equation}
    \begin{split}
        &I_B^{(n)}\sim\e^{-\frac{\lb}{4} \left(2\left(k^y-(k')^y\right)^2 + \left( k^x -(k')^x +\iu \left( k^y - (k')^y \right) \right)^2\right)} \hspace{-0.4em}\int_0^\infty \hspace{-0.4em}\dd{r} r^2 \e^{-\frac{r^2}{2\lb}} L_n^1\left(\frac{r^2}{2\lb}\right) \\
        &\cross\int_0^{2\pi} \dd{\varpi} \left((\Vec{\gamma}_\perp\vdot\Hat{\upsilon})  \cos{\varpi} + (\Vec{\gamma}_\perp\vdot\Hat{\upsilon}_\perp) \sin{\varpi} \right) \e^{\iu r\upsilon \cos{\varpi}} \\
        &= -2\pi\lambda_B^4 \iu (\Vec{\upsilon}\vdot \Vec{\gamma}_\perp) \e^{-\frac{1}{4}\lb\left(k_\perp^2 + (k')_\perp^2\right)}\e^{\iu\Theta}\e^{\iu\alpha}\frac{\left( \lb k_\perp k_\perp'\e^{\iu(\phi'-\phi)} \right)^n}{2^n n!},
    \end{split}
\end{equation}
where $\Hat{\upsilon}_\perp$ is a unit vector perpendicular to $\Hat{\upsilon}$, and we have used the known integrals \citep[eq. 7.421.4]{gradshtein_table_2015}
% \vspace{2em}
\begin{equation}\label{eq:GenLaguerreInt}
\begin{split}
    &\int_0^\infty \dd{x} x^{\nu+1}\e^{-\beta x^2} L_n^\nu (\vartheta x^2) J_\nu (xy) \\
    &= 2^{-\nu-1}\beta^{-\nu-n-1}(\beta-\vartheta)^n y^\nu \e^{-\frac{y^2}{4\beta}} L_n^\nu \left( \frac{\vartheta y^2}{4\beta(\vartheta-\beta)} \right),
\end{split}
\end{equation}
\begin{equation}\label{eq:BesselIntCos}
    \int_0^{2\pi} \dd{\phi}\cos{\phi}\e^{\iu x\cos{\phi}}=2\pi\iu J_1(x),
\end{equation}
and
\begin{equation}\label{eq:BesselIntSin}
    \int_0^{2\pi} \dd{\phi}\sin{\phi}\e^{\iu x\cos{\phi}}=0.
\end{equation}
with $\vartheta~=~\beta~=~\flatfrac{1}{2}$, recalling the leading behavior of the generalized Laguerre polynomials $L_n^\alpha(x)\sim (-1)^n \flatfrac{x^n}{n!}$. Thus, the original integral yields
\begin{equation}
    \begin{split}
        I_B^{(n)}&=-(2\pi)^3\iu e^2\delta^{(3)}(E,p^z,p^y) \e^{-\frac{1}{4}\lb\left(k_\perp^2 + (k')_\perp^2\right)} \e^{\iu \Theta} \e^{\iu \alpha}\\
        &\cross\Bar{B}(\Vec{\upsilon}\vdot \Vec{\gamma}_\perp)B \frac{\left( \lb k_\perp k_\perp'\e^{\iu(\phi'-\phi)} \right)^n}{2^n n!}.
    \end{split}
\end{equation}
As expected, performing the same calculation for the $u$-channel yields the same result with the substitutions $k \leftrightarrow -k'$ and $\varepsilon \leftrightarrow \varepsilon^*$, i.e.,~$\Theta \leftrightarrow -\Theta$ and $\Delta\phi \leftrightarrow -\Delta\phi$.

Combining the results obtained above, the scattering amplitudes read
\begin{widetext}
\begin{equation}\label{eq:ScatteringAmplitudes}
    \begin{split}
        \iu T_s^{(n)} &= -\iu e^2 (2\pi)^3\delta^{(3)}(E,p^z,p^y)\frac{\e^{-\frac{1}{4}\lb\left(k_\perp^2 + (k')_\perp^2\right)} \e^{\iu \Theta} \e^{\iu \alpha} \e^{\iu n\Delta\phi}}{E_I^2-\left(p_I^z\right)^2-m_e^2-2eBn}\frac{\left(\lb k_\perp k_\perp'\right)^n}{\sqrt{E'+m_e} 2^n n!}\sqrt{2m_e} \Bigg\{f_i^z(\theta)f_{i'}^z(\theta')\left((E_I-m_e)(E'+m_e) + p_I^z (p')^z\right) \\
        &+\frac{2n}{\lb k_\perp k_\perp'}\bigg[f_i^-(\theta)(f_{i'}^-(\theta'))^*\left((E_I-m_e)(E'+m_e) - p_I^z (p')^z\right) - (p')^z\left(k'_\perp f_i^-(\theta)f_{i'}^z(\theta') + k_\perp f_i^z(\theta) (f_{i'}^-(\theta'))^*\right)\bigg]\Bigg\} \\
        & \equiv -\frac{\iu e^2 (2\pi)^3\sqrt{2m_e}}{\sqrt{E'+m_e}}\delta^{(3)}(E,p^z,p^y)\e^{-\frac{1}{4}\lb\left(k_\perp^2 + (k')_\perp^2\right)} \e^{\iu \Theta} \e^{\iu \alpha} \e^{\iu n\Delta\phi} F_+^{(n)},\\
        \iu T_u^{(n)} & = -\iu e^2 (2\pi)^3\delta^{(3)}(E,p^z,p^y)\frac{\e^{-\frac{1}{4}\lb\left(k_\perp^2 + (k')_\perp^2\right)} \e^{-\iu \Theta} \e^{\iu \alpha} \e^{-\iu n\Delta\phi}}{E_I^2-\left(p_I^z\right)^2-m_e^2-2eBn}\frac{\left(\lb k_\perp k_\perp'\right)^n}{\sqrt{E'+m_e} 2^n n!}\sqrt{2m_e}\Bigg\{f_i^z(\theta)f_{i'}^z(\theta')\left((E_I-m_e)(E'+m_e) + p_I^z (p')^z\right) \\
        &+\frac{2n}{\lb k_\perp k_\perp'}\bigg[f_i^+(\theta)(f_{i'}^+(\theta'))^*\left((E_I-m_e)(E'+m_e) - p_I^z (p')^z\right) + (p')^z\left(k'_\perp f_i^+(\theta)f_{i'}^z(\theta') + k_\perp f_i^z(\theta) (f_{i'}^+(\theta'))^*\right)\bigg]\Bigg\}\\
        & \equiv -\frac{\iu e^2 (2\pi)^3\sqrt{2m_e}}{\sqrt{E'+m_e}}\delta^{(3)}(E,p^z,p^y)\e^{-\frac{1}{4}\lb\left(k_\perp^2 + (k')_\perp^2\right)} \e^{-\iu \Theta} \e^{\iu \alpha} \e^{-\iu n\Delta\phi} F_-^{(n)},
    \end{split}
\end{equation}
where $i, i' \in$ \{O,X\}. Finally, we obtain the differential cross section
\begin{equation}\label{eq:FinalCrossSec}
    \begin{split}
        \dv{\sigma}{\cos{\theta'}} &= \frac{1}{2} \frac{\omega'}{\omega}\frac{2\pi r_e^2 m_e^2 \e^{-\frac{1}{2}\lb\left(k_\perp^2 + (k')_\perp^2\right)}}{(E'+m_e)(E'-(p')^z \cos{\theta'})} \Bigg[ \sum_{n=0}^\infty \sum_{\nu\in\{+,-\}} \abs{F_\nu^{(n)}}^2+ 2\sum_{n,n'=0}^\infty \Re{F_+^{(n)} \left( F_-^{(n')} \right)^*} J_{n+n'}\left(\lb \omega\omega'\sin{\theta}\sin{\theta'}\right) \Bigg].
    \end{split}
\end{equation}
\end{widetext}
In the calculation, we have used the integrals
\begin{equation}
    \int_0^{2\pi} \dd{\Delta\phi} \e^{\iu\nu(n-n')\Delta\phi} = 0,\quad \nu=\nu',
\end{equation}
and
\begin{equation}
    \begin{split}
        &\int_0^{2\pi} \dd{\Delta\phi} \e^{-\iu\nu \lb \omega\omega' \sin{\theta}\sin{\theta'}\sin(\Delta\phi) + \nu(n+n')\Delta\phi} \\
        &= 2\pi J_{n+n'}\left(\lb \omega\omega' \sin{\theta}\sin{\theta'}\right),\quad \nu=-\nu'.
    \end{split}
\end{equation}

We have performed the calculation of the Compton scattering cross section using the Landau-level projection of the Schwinger proper time propagator. This is a reasonable choice because the terms in the obtained cross section are suppressed for large $n$ \citep{kozlenkov_two-photon_1986}. The same calculation can also be done using the the propagator in its Schwinger proper time form. For completeness, this calculation is shown in App.~\ref{app:ComptonSchwinger}. The Landau level projected result is preferred in most applications due to the direct access to the individual Landau levels.

The final phase of calculating the cross section of Compton scattering is to also take into account the excited Landau levels of the external electrons. In the following calculations, the Landau levels are not restricted in any way. However, not all Landau levels are accessible for the electrons. Incoming electrons in a strong magnetic field rapidly decay into the LLL due to synchrotron radiation, while the Landau level of the outgoing electron is limited by the total energy of the incoming particles such that \citep{mushtukov_compton_2016}
\begin{equation}
    n_\mathrm{f,max}=n_\mathrm{i}+\left\lfloor \frac{k_\mathrm{R}^2}{2bm^2} \right\rfloor,
\end{equation}
where
\begin{equation}
    k_\mathrm{R}^2 \equiv \omega^2\sin[2](\theta)+2\omega(E-p^z\cos{\theta}) = k_\parallel^2+2k_\parallel\vdot p_\parallel.
\end{equation}

To proceed, we note that
\begin{equation}
\begin{split}
    -\iu\dv{k_x}\e^{-\iu k\vdot x} &= x \e^{-\iu k\vdot x}, \quad \iu\dv{(k_x)'}\e^{\iu k'\vdot x'} = x' \e^{\iu k'\vdot x'}, \\
    -\iu\dv{k_x}\e^{-\iu k\vdot x'} &= x' \e^{-\iu k\vdot x'}, \quad \iu\dv{(k_x)'}\e^{\iu k'\vdot x} = x \e^{\iu k'\vdot x}.
\end{split}
\end{equation}
Thus, we can evaluate the scattering amplitude integrals by making the substitutions
\begin{equation}
\begin{split}
        H_{n_\mathrm{i}}&\left(\frac{x-a}{\lambda_B}\right)H_{n_\mathrm{f}}\left(\frac{x'-a'}{\lambda_B}\right)\\
        \to H_{n_\mathrm{i}}&\left(\frac{-\iu\dv{k_x}-a}{\lambda_B}\right)H_{n_\mathrm{f}}\left(\frac{\iu\dv{(k_x)'}-a'}{\lambda_B}\right)
\end{split}
\end{equation}
and 
\begin{equation}
\begin{split}
    H_{n_\mathrm{i}}&\left(\frac{x-a}{\lambda_B}\right)H_{n_\mathrm{f}}\left(\frac{x'-a'}{\lambda_B}\right) \\
    \to H_{n_\mathrm{i}}&\left(\frac{\iu\dv{(k_x)'}-a}{\lambda_B}\right)H_{n_\mathrm{f}}\left(\frac{-\iu\dv{k_x}-a'}{\lambda_B}\right)
\end{split}
\end{equation}
for the $s$ and $u$-channels, respectively. Here, it is vital to demand that the derivatives do not act on the polarization vectors. All terms in the scattering amplitudes are proportional to
\begin{equation}
    A_{a,b;0,0}^{(s)}\equiv\e^{-\frac{1}{4}\lb\left(k_\perp^2 + (k')_\perp^2\right)} \e^{\iu \Theta} \e^{\iu \alpha}\left(k_\perp\e^{-\iu\phi}\right)^a \left(k_\perp'\e^{\iu\phi'}\right)^b,
\end{equation}
or
\begin{equation}
    A_{a,b;0,0}^{(u)}\equiv\e^{-\frac{1}{4}\lb\left(k_\perp^2 + (k')_\perp^2\right)} \e^{-\iu \Theta} \e^{\iu \alpha}\left(k_\perp\e^{\iu\phi}\right)^a \left(k_\perp'\e^{-\iu\phi'}\right)^b,
\end{equation}
where $a,b \in \{n-1,n\}$.

We make the ansatz
\begin{widetext}
\begin{equation}\label{eq:AsGen}
\begin{split}
    A_{a,b;n_\mathrm{i},n_\mathrm{f}}^{(s)}&\equiv n_\mathrm{i}!n_\mathrm{f}!\left(\frac{-2\iu}{k_\perp\lambda_B}\right)^{n_\mathrm{i}}\left(\frac{2\iu}{k_\perp'\lambda_B}\right)^{n_\mathrm{f}} L_{n_\mathrm{i}}^{a-n_\mathrm{i}}\left(\frac{k_\perp^2\lb}{2}\right)L_{n_\mathrm{f}}^{b-n_\mathrm{f}}\left(\frac{(k_\perp')^2\lb}{2}\right)
    \exp(\iu\left(\frac{n_\mathrm{i}-n_\mathrm{f}}{2}(\phi'+\phi)-\frac{n_\mathrm{i}+n_\mathrm{f}}{2}(\phi'-\phi)\right))A_{a,b;0,0}^{(s)},
\end{split}
\end{equation}
and
\begin{equation}\label{eq:AuGen}
\begin{split}
    A_{a,b;n_\mathrm{i},n_\mathrm{f}}^{(u)}&\equiv n_\mathrm{i}!n_\mathrm{f}!\left(\frac{-2\iu}{k_\perp\lambda_B}\right)^{n_\mathrm{f}}\left(\frac{2\iu}{k_\perp'\lambda_B}\right)^{n_\mathrm{i}} L_{n_\mathrm{f}}^{a-n_\mathrm{f}}\left(\frac{k_\perp^2\lb}{2}\right)L_{n_\mathrm{i}}^{b-n_\mathrm{i}}\left(\frac{(k_\perp')^2\lb}{2}\right)
    \exp(\iu\left(\frac{n_\mathrm{i}-n_\mathrm{f}}{2}(\phi'+\phi)+\frac{n_\mathrm{i}+n_\mathrm{f}}{2}(\phi'-\phi)\right))A_{a,b;0,0}^{(u)},
\end{split}
\end{equation}
\end{widetext}
We will prove the ansatz by induction. It is straightforward to check that the ansatz is correct for any combination of $n_\mathrm{i},n_\mathrm{f} \in \{0,1\}$. In the following, we will focus on the case $n_\mathrm{f}=0$ for the sake of clarity. The results can be generalized to non-zero $n_\mathrm{f}$ without any extra work since the $n_\mathrm{i}$ and $n_\mathrm{f}$ contributions are independent of each other. The Hermite polynomials satisfy the recursion relation
\begin{equation}
    H_{n+1}(x) = 2xH_n(x) - 2nH_{n-1}(x) = H_1(x)H_n(x) - 2nH_{n-1}(x).
\end{equation}
Thus, we find that
\begin{equation}
\begin{split}
    &H_{n_\mathrm{i}+1}\left(\frac{-\iu\dv{k_x}+\lb p_y}{\lambda_B}\right)A_{a,b;0,0}^{(s)} \\
    &=  2\left(\frac{-\iu\dv{k_x}+\lb p_y}{\lambda_B}\right)H_{n_\mathrm{i}}\left(\frac{-\iu\dv{k_x}+\lb p_y}{\lambda_B}\right)A_{a,b;0,0}^{(s)} \\
    &- 2 n_\mathrm{i} H_{n_\mathrm{i}-1}\left(\frac{-\iu\dv{k_x}+\lb p_y}{\lambda_B}\right) A_{a,b;0,0}^{(s)}\\
    &= 2\left(\frac{-\iu\dv{k_x}+\lb p_y}{\lambda_B}\right) A_{a,b;n_\mathrm{i},0}^{(s)} - 2n_\mathrm{i} A_{a,b;n_\mathrm{i}-1,0}^{(s)}.
\end{split}
\end{equation}
We rewrite the derivative with respect to $k_x$ with the substitution
\begin{equation}
    \Xi = \frac{\lb k_\perp^2}{2}.
\end{equation}
Then, the derivative reads
\begin{equation}
    \pdv{k_x} = \lambda_B\left(\sqrt{2\Xi}\left(\frac{\e^{\iu\phi}+\e^{-\iu\phi}}{2}\right)\pdv{\Xi}-\frac{1}{\sqrt{2\Xi}}\left(\frac{\e^{\iu\phi}-\e^{-\iu\phi}}{2\iu}\right)\pdv{\phi}\right).
\end{equation}
Acting with the derivative and using the recursion relation of the generalized Laguerre polynomials
\begin{equation}
    L^\alpha_n(x) = \frac{\alpha+1-x}{n}L^{\alpha+1}_{n-1}(x) -\frac{x}{n}L^{\alpha+2}_{n-2}(x),
\end{equation}
the result
\begin{equation}
    H_{n_\mathrm{i}+1}\left(\frac{-\iu\dv{k_x}+\lb p_y}{\lambda_B}\right)A_{a,b;0,0}^{(s)} = A_{a,b;n_\mathrm{i}+1,0}^{(s)}
\end{equation}
can be established. Thus, since the ansatz is true for $n_\mathrm{i}=0$ and $n_\mathrm{i} = 1$, and ${A_{a,b;n_\mathrm{i},0}^{(s)} \Rightarrow A_{a,b;n_\mathrm{i}+1,0}^{(s)}}$, the ansatz is true for all $n_\mathrm{i},n_\mathrm{f}$ by induction. The full ansatz for $A_{a,b;n_\mathrm{i},n_\mathrm{f}}^{(s)}$ and $A_{a,b;n_\mathrm{i},n_\mathrm{f}}^{(u)}$ can be proven in the same way. We also note that the extra phase shifts caused by the excited Landau levels of the electron legs can be absorbed into the definitions of $\Theta$ and $\alpha$ given in Eq. \eqref{eq:ThetaAlpha}:
\begin{equation}
\begin{split}
    \Theta &= -\frac{1}{2}\lb k_\perp k'_\perp\sin(\Delta\phi) + \left( n - \frac{n_\mathrm{i}+n_\mathrm{f}}{2}\right)\Delta\phi, \\
    \alpha &= \frac{1}{2} \lb \left(k_x'-k_x\right)\left(p_y+p_y'\right) + \frac{n_\mathrm{i}-n_\mathrm{f}}{2}(\phi+\phi').
\end{split}
\end{equation}
These same phases were also found in Ref.~\cite{mushtukov_compton_2016}. The final scattering amplitudes for Compton scattering are given in App.~\ref{app:ComptonResult}.

No full analytical formulas for the scattering amplitudes were provided in Ref.~\citep{mushtukov_compton_2016}, and thus, a direct comparison of the obtained expressions is not possible. Instead, I have compared the scattering amplitudes with the ones provided in Ref.~\citep{daugherty_compton_1986}, where the calculation has been done using JL wave functions for the electrons and with the incoming electron on the LLL. Taking these differences into account, I find that the two results match except for a few sign errors.%
\footnote{%
In eq.~(8a) of Ref.~\citep{daugherty_compton_1986}, there is a missing minus sign in front of the terms proportional to $\Lambda_{n_\mathrm{f},n}(\beta_f)\Lambda_{n-1,0}(\beta_i)$, $\Lambda_{n_\mathrm{f},n-1}(\beta_f)\Lambda_{n,0}(\beta_i)$, $\Lambda_{n_\mathrm{f}-1,n}(\beta_f)\Lambda_{n,0}(\beta_i)$, and $\Lambda_{n_\mathrm{f}-1,n-1}(\beta_f)\Lambda_{n-1,0}(\beta_i)$. Additionally, there is an error in the overall normalization factor in their eq.~(11), where $m+\omega+\omega'=E'$ in the denominator should be replaced by $E'-(p')^z\cos{\theta'}$.}

\subsection{Two-photon pair creation}
The cross section of two-photon pair creation is obtained from the cross section of Compton scattering by the substitutions
\begin{equation}
    \begin{split}
        &p^\mu\to -p^\mu,\quad (k')^\mu \to -(k')^\mu \\
        &\us{n}{a}{\sigma}{x} \to \vs{n}{a}{\sigma}{x},\quad \varepsilon_{k'} \to \varepsilon_{k'}^*.
    \end{split}
\end{equation}
We choose to do the calculation in the Co$z$M frame. To calculate the cross section, we need to know the relative velocity of the two incoming photons, which is given by Eq.~\eqref{eq:MollerFlux} and reads
\begin{equation}
\begin{split}
    \Bar{v} &= \sqrt{(\Vec{v}-\Vec{v}')^2 - \left(\Vec{v}\cross\Vec{v}'\right)^2} = 1-\cos{\theta_{12}}.
\end{split}
\end{equation}
Here, $\theta_{12}$ is the angle between the two photons. The total cross section reads
\begin{equation}
    \begin{split}
        \sigma_\mathrm{2PC} &= \frac{1}{2\pi\Bar{v}} \int \frac{\dd[2]{p'}}{2E'}\frac{\dd[2]{p}}{2E}\delta^{(3)}(E,p^z,p^y)\frac{\abs{M_\mathrm{2PC}}^2}{4\omega\omega'} \\
        &= \sum_{p^z}\frac{1}{2\pi\lb \Bar{v}} \abs{\frac{p^z}{E} - \frac{(p')^z}{E'}}^{-1}\frac{\abs{M_\mathrm{2PC}}^2}{16EE'\omega\omega'}.
    \end{split}
\end{equation}
Note that compared to Compton scattering, we do not integrate over any angles, and thus, the result depends on $\theta$, $\theta'$ and $\Delta \phi$. Like in the one-photon case, the two-photon pair creation cross section is amplified by a factor of $b$ in strong magnetic fields compared to the Compton scattering cross section.

\subsection{Two-photon pair annihilation}
The cross section of two-photon pair annihilation is obtained from the cross section of Compton scattering by the substitutions
\begin{equation}
\begin{split}
    &(p')^\mu\to -(p')^\mu,\quad k^\mu \to -k^\mu, \\
    &\ubs{n'}{a'}{\sigma'}{x'} \to \vbs{n'}{a'}{\sigma'}{x'},\quad \varepsilon_{k} \to \varepsilon_{k}^*.
\end{split}
\end{equation}
We perform the calculation in the Co$z$M frame, but the rest frame of either fermion might be preferred in some applications. This is, however, not a problem since the cross sections are invariant under boosts along the magnetic field. Since the two outgoing photons are identical we must divide the final result by two \cite{peskin_introduction_1995}.
The total cross section reads
\begin{equation}
    \begin{split}
        \sigma_\mathrm{2PA} &= \frac{\lambda_B^4}{2\Bar{v}}\int\hspace{-0.25em}\dd{p^y}\dd{(p')^y}\hspace{-0.5em} \int\frac{\dd[3]{k}}{2\omega(2\pi)^3}\frac{\dd[3]{k'}}{2\omega'}\delta^{(3)}(E,p^z,p^y)\frac{\abs{M_\mathrm{2PA}}^2}{4EE'} \\
        &= \frac{\lb}{(2\pi)^3 2\Bar{v}} \int\dd{\Omega}\dd{\Omega'} \frac{\omega \omega'}{\abs{\cos{\theta} - \cos{\theta'}}} \frac{\abs{M_\mathrm{2PA}}^2}{16EE'}.
    \end{split}
\end{equation}
The integration over two solid angles requires a careful inspection of the integration bounds. See App. \ref{app:DoubleIntegration} for more details. The cross section of two-photon pair annihilation is suppressed by a factor of $b^{-1}$ in strong magnetic fields compared to the Compton scattering cross section.

\subsection{\label{sec:otherAlphaTwo}Other $\mathcal{O}(\alpha_e^2)$ cross sections}
There are also other QED scattering processes with $\mathcal{O}(\alpha_e^2)$ cross sections that could be relevant for the plasma dynamics of magnetar magnetospheres. M\o ller and Bhabha scattering have a photon propagator instead of a fermion propagator and must therefore be treated separately. The calculation of their cross sections follow the same recipe presented in this paper and will be discussed in a follow-up paper. However, these two scattering processes are not expected to play a key role in the plasma dynamics of magnetar magnetospheres because the number density of $e^\pm$ is much lower than for photons in the magnetosphere. Thus, scatterings that require two $e^\pm$ to collide are suppressed compared to ones that include a photon. 

In addition to 2-to-2-particle scattering processes, there exists also another category of scattering processes with $\mathcal{O}(\alpha_e^2)$ cross sections that might be relevant to magnetars---1-to-3-particle scattering processes. They can be thought of as higher-order corrections to 1-to-2-particle processes and are expected to enhance, e.g., $\Gamma_\mathrm{SR}$ when the outgoing photon is near the pair-creation threshold. The calculation of 1-to-3-particle decay processes can be done via crossing symmetry from the cross section of Compton scattering.

\subsection{\label{sec:higherOrder}Higher-order processes}
There is one more 3-particle process that has been suggested to be relevant for the plasma dynamics of magnetar magnetospheres---photon splitting \citep{beloborodov_corona_2007, harding_pair_2025}. In ultra-strong magnetic fields most electrons and positrons are expected to be trapped on the lowest Landau level, which makes synchrotron radiation ineffective for producing particle cascades. Similarly, one-photon-pair creation is only activated when the energy of the photon is at least $2m_e$ and thus low-energy photons cannot produce particle cascades through pair creation. Neither of these issues are present for photon splitting, which suggests that photon splitting could play a vital role in low-energy particle cascades in magnetar magnetospheres. 

Unlike the other 3-particle scattering processes, which have cross sections of $\mathcal{O}(\alpha_e)$, the photon splitting-cross section is of $\mathcal{O}(\alpha_e^3)$ and its calculation is more difficult because the Feynman diagram includes a fermion loop. The cross section was first calculated in Ref.~\citep{adler_photon_1970}. In a follow-up paper, we will show how our new formalism can be used to calculate loops in SFQED.

Another group of interesting scattering processes with cross sections of $\mathcal{O}(\alpha_e^3)$ are 2-to-3-particle scatterings. They are higher-order corrections to 2-to-2-particle scattering processes that might be relevant for example if an outgoing photon is near the threshold for pair creation. 
Photon-producing processes (such as Double Compton $e^\pm + \gamma \to e^\pm + \gamma + \gamma$) are also important during pair cascades because they do not conserve the particle number during the interaction and can therefore drive photon distributions from Wien-like into a Planck distribution \citep{nattila_radiative_2024}.
The cross sections of these higher-order scattering processes can be calculated with the same methods already discussed in this paper. The key steps of the calculation are to use CoM-coordinates, start with the LLLa calculation, and then generalize to excited Landau levels with known integrals and recursion relations.

\section{\label{sec:summary}Summary}

In this article, I have presented a systematic formalism for calculating the cross sections of QED scattering processes in a strong background magnetic field. To demonstrate the effectiveness of the formalism, I have calculated the cross sections of all tree-level 1-to-2, 2-to-1, and 2-to-2 particle QED scattering processes that do not include a photon propagator. These are the main scattering processes that are expected to be relevant for the magnetospheric plasma dynamics of magnetars. The obtained cross sections are compared with earlier results in the companion letter \citep{kiuru_qed_2026} and made available through an open-source Python package\footnotemark[1]. The companion letter also contains a more in-depth discussion on the physical context and impact of these results.

 The formalism used in this paper is straightforward to generalize to the calculation of higher-order scattering processes, due to the systematic nature of the formalism. However, important questions still remain about the validity of perturbation theory in the calculation of loop correction in SFQED. One especially interesting future avenue of research would be to investigate the validity of the Ritus-Narozhny conjecture of laser plasmas (where the background field is an electromagnetic plane wave \citep{fedotov_advances_2023}) in the context of constant background magnetic fields.

\begin{acknowledgments}
The author is grateful to Joonas Nättilä, Risto Paatelainen, and Aleksi Vuorinen for insightful discussions and mentorship. This work was supported by the Research Council of Finland, the Centre of Excellence in Neutron-Star Physics (projects 374062 and 374063), ERC grant (ILLUMINATOR, 101114623), and the Finnish Cultural Foundation.
\end{acknowledgments}

\bibliography{apssamp}% Produces the bibliography via BibTeX.

\appendix
\section{\label{app:ElecProp}Derivation of the electron propagator}

In this appendix, I derive the electron propagator in an external magnetic field. The propagator is first derived with the Landau levels explicitly shown. Subsequently, the Schwinger proper time formalism is used to remove the explicit dependency on the Landau levels. The derivation mostly follows the one presented in Ref.~\citep{shovkovy_magnetic_2013}.

The position space propagator is formally given by the expression
\begin{equation}
    S_\mathrm{F}(x,x')\equiv\iu\left(\slashed{p}-m_e\right)^{-1}\delta^{(4)}(x-x'),
\end{equation}
where $p_\mu=\iu\partial_\mu-qA_\mu$. It is most convenient to work in a mixed representation, where a Fourier transformation in the variables $t-t'$ and $z-z'$ has been done. This yields
\begin{widetext}
\begin{equation}
\label{eq:AppPropFourier}
    S_\mathrm{F}(E,p^z;\Vec{x}_\perp,\Vec{x}'_\perp)=\iu\left(\g{0}E-\g{3}p^z-\g{1}p^x-\g{2}p^y-m_e\right)^{-1}\delta^{(2)}(\Vec{x}_\perp-\Vec{x}_\perp').
\end{equation}
Making use of the identity
\begin{equation}
    \left(\slashed{a}-m_e\right)\left(\slashed{a}+m_e\right)=\slashed{a}\slashed{a}-m_e^2,
\end{equation}
Eq.~\ref{eq:AppPropFourier} can be written in the form
\begin{equation}
    S_\mathrm{F}(E,p^z;\Vec{x}_\perp,\Vec{x}'_\perp) =\iu\left(\g{0}E-\g{3}p^z-\g{1}p^x-\g{2}p^y+m_e\right) \left[E^2-(p^x)^2 - (p^y)^2 + \iu \abs{qB}\g{1}\g{2} - (p^z)^2 - m_e^2 \right]^{-1}\delta^{(2)}(\Vec{x}_\perp-\Vec{x}_\perp').
\end{equation}

The propagator can be further simplified by making use of the completeness relations of the Hermite functions, given in Eq.~\eqref{eq:HermiteComp}, to replace the Dirac delta function, as well as the fact that the Hermite functions are eigenstates of the $\Vec{p}_\perp^2$ operator with eigenvalue $(2k+1)/\lb$. With these simplifications, the propagator reads
\begin{equation}
    S_\mathrm{F}(E,p^z;\Vec{x}_\perp,\Vec{x}'_\perp)=\iu\int\dd{p^y}\sum_{k=0}^\infty\frac{\g{0}E-\g{3}p^z-\g{1}p^x-\g{2}p^y+m}{E^2 - \frac{2k+1}{\lb} + \frac{\iu s_\pm\g{1}\g{2}}{\lb} - (p^z)^2 - m^2} \Psi_{k,a}(\Vec{x}_\perp)\Psi_{k,a}^*(\Vec{x}'_\perp)
\end{equation}
The Hermite function recursion relations
\begin{equation}
    \phi_{n,a}'(\varsigma)=\sqrt{\frac{n}{2}}\phi_{n-1,a}(\varsigma)-\sqrt{\frac{n+1}{2}}\phi_{n+1,a}(\varsigma), \quad     \varsigma\phi_{n,a}(\varsigma)=\sqrt{\frac{n}{2}}\phi_{n-1,a}(\varsigma)+\sqrt{\frac{n+1}{2}}\phi_{n+1,a}(\varsigma),
\end{equation}
where $'\equiv\dv*{\varsigma}$ and $\dv*{\varsigma}=\lambda_B\dv*{x}$, yield the relation
\begin{equation}
    \left(p^x\g{1}+p^y\g{2}\right)\Psi_{k,a}=\frac{\iu}{\lambda_B}\g{1}\left(\sqrt{2(k+1)}\mathcal{P}_{-} \Psi_{k+1,a} - \sqrt{2k}\mathcal{P}_{+} \Psi_{k-1,a} \right).
\end{equation}
Thus, the propagator reads
\begin{equation}\label{eq:PropLandau}
    S_\mathrm{F}(E,p^z;\Vec{x}_\perp,\Vec{x}_\perp')=\iu\sum_{k=0}^\infty \frac{AI_1^{(k)} + BI_2^{(k)} + CI_3^{(k)}}{E^2 - \frac{2k+1}{\lb} + \iu \abs{qB}\g{1}\g{2} - (p^z)^2 - m_e^2},
\end{equation}
where
\begin{equation}
\begin{split}
    A &= \frac{\iu\g{1}}{\lambda_B}\sqrt{2k}\mathcal{P}_{+},\quad B = \g{0}E-\g{3}p^z+m_e,\quad C = -\frac{\iu\g{1}}{\lambda_B}\sqrt{2(k+1)}\mathcal{P}_{-},\\
    I_1^{(k)} &= \int\dd{p^y}\Psi_{k-1,a}\Psi^*_{k,a}(\Vec{x}'_\perp),\quad I_2^{(k)} = \int\dd{p^y}\Psi_{k,a}(\Vec{x}_\perp)\Psi^*_{k,a}(\Vec{x}'_\perp),\quad
    I_3^{(k)} = \int\dd{p^y}\Psi_{k+1,a}\Psi^*_{k,a}(\Vec{x}'_\perp).
\end{split}
\end{equation}

There are three integrals to evaluate. The second integral reads
\begin{equation}
\begin{split}
    I_2^{(k)}&=\int\dd{p^y}\Psi_{k,a}(\Vec{x}_\perp)\Psi^*_{k,a}(\Vec{x}'_\perp)
    = N^2 \int \dd{p^y} H_k\left(\frac{x}{\lambda_B}-s_\pm p^y\lambda_B\right)H_k\left(\frac{x'}{\lambda_B}-s_\pm p^y\lambda_B\right)\e^{-\frac{\left(\frac{x}{\lambda_B}-s_\pm p^y\lambda_B\right)^2}{2}}\e^{-\frac{\left(\frac{x'}{\lambda_B}-s_\pm p^y\lambda_B\right)^2}{2}}\e^{\iu p^y(y-y')},
\end{split}
\end{equation}
where $N$ is a normalization constant. The integral is easiest to evaluate with the substitution
\begin{equation}
    u = -s_\pm p^y \lambda_B+\frac{x+x'}{2\lambda_B}.
\end{equation}
\begin{equation}
\begin{split}
    I_2^{(k)}&=\frac{N^2}{\lambda_B} \int \dd{u} H_k\left( u + \frac{x-x'}{2\lambda_B} \right) H_k\left( u + \frac{x'-x}{2\lambda_B} \right)\e^{-\frac{1}{2}\left[\left( u + \frac{x-x'}{2\lambda_B} \right)^2 + \left( u + \frac{x'-x}{2\lambda_B} \right)^2 \right]}\e^{-\iu s_\pm \left(\frac{u}{\lambda_B} - \frac{x+x'}{2\lb}\right)(y-y')}.
\end{split}
\end{equation}
The final exponential function gives the Schwinger phase $\e^{i\Phi}$, where $\Phi=\frac{(x+x')(y-y')}{2\lb}$. With the substitutions
\begin{equation}
    \begin{split}
    v &\equiv u + \frac{\iu s_\pm}{2\lambda_B}(y-y') \equiv u + b,\quad \xi^2\equiv\frac{(x-x')^2 + (y-y')^2}{2\lb}
    \end{split}
\end{equation}
the integral reads
\begin{equation}
    I_2^{(k)}=\frac{N^2\e^{\iu\Phi-\frac{\xi^2}{2}}}{\lambda_B} \int \dd{v} H_k\left( v + \frac{x-x'}{2\lambda_B} - b \right) H_k\left( v + \frac{x'-x}{2\lambda_B} - b \right)\e^{-v^2},
\end{equation}
which is a known integral with the result \citep{gradshtein_table_2015}
\begin{equation}\label{eq:HermiteInt}
    \int \dd{x} H_m(x+y) H_n(x+z) \e^{-x^2} = 2^n \pi^\frac{1}{2}m! z^{n-m} L_m^{n-m}(-2yz),\quad n\geq m.
\end{equation}
In our case, $y=\frac{x-x'}{2\lambda_B}-b$ and $z=\frac{x'-x}{2\lambda_B} - b$ and, therefore, $-2yz=\xi^2$. Finally, the integral evaluates to
\begin{equation}
    \sum_{k=0}^\infty I_2^{(k)} = \frac{1}{2\pi\lb}\e^{\iu\Phi}\e^{-\frac{\xi^2}{2}}\sum_{k=0}^\infty L_k(\xi^2). %= \frac{1}{2\pi\lb}\e^{\iu\Phi}\e^{-\frac{\xi^2}{2}}\sum_{k=0}^\infty \left(\mathcal{P}_{-} +\mathcal{P}_{+}\right)L_k(\xi^2).
\end{equation}
The summation index is changed from $k$ to the Landau level index
\begin{equation}
    n = k -s_\pm\frac{\mu}{2}+\frac{1}{2},
\end{equation}
where $\mu=\pm 1$ is the spin state of the propagator, with the identity
\begin{equation}
\begin{split}
    &\frac{1}{E^2 - (2k+1)\abs{qB} + \iu s_\pm\abs{qB}\g{1}\g{2} - (p^z)^2 - m_e^2} = \frac{E^2 - 2(k+\mathcal{P}_+)\abs{qB} - (p^z)^2 - m_e^2}{(E^2 - 2(k+1)\abs{qB} - (p^z)^2 - m_e^2)(E^2 - 2k\abs{qB} - (p^z)^2 - m_e^2)}.
\end{split}
\end{equation}
The result reads
\begin{equation}
\begin{split}
    &\sum_{k=0}^\infty \frac{B I_2^{(k)}}{E^2 - (2k+1)\abs{qB} + \iu s_\pm\abs{qB}\g{1}\g{2} - (p^z)^2 - m_e^2} = \frac{\e^{\iu\Phi}\e^{-\frac{\xi^2}{2}}}{2\pi\lb} \sum_{n=0}^\infty \frac{\left(\g{0}E-\g{3}p^z+m_e\right)\left( \mathcal{P}_{-}L_{n-1}(\xi^2) + \mathcal{P}_{+}L_n(\xi^2) \right)}{E^2-(p^z)^2-m_e^2-2\abs{qB}n},
\end{split}
\end{equation}
where the projection operator properties
\begin{equation}
    \mathcal{P}_\pm^2=\mathcal{P}_\pm,\quad \mathcal{P}_+\mathcal{P}_-=0,\quad \mathcal{P}_+ + \mathcal{P}_- = 1,
\end{equation}
have been used. The integrals $I_1^{(k)}$ and $I_3^{(k)}$ can be evaluated in a similar manner. In the end, the propagator in the mixed representation reads
\begin{equation}
\begin{split} \label{eq:PropLevelsApp}
    &S_\mathrm{F}(E,p^z;\Vec{x}_\perp,\Vec{x}'_\perp)=\iu \frac{\e^{\iu\Phi}\e^{-\frac{\xi^2}{2}}}{2\pi\lb} \sum_{n=0}^\infty \frac{F_n}{E^2-(p^z)^2-m_e^2-2n\abs{qB}}, \\
    F_n&=\left(\g{0}E-\g{3}p^z+m_e\right)\left( \mathcal{P}_{-}L_{n-1}(\xi^2) + \mathcal{P}_{+}L_n(\xi^2) \right) - \frac{\iu}{\lb} \Vec{\gamma}_\perp\vdot(\Vec{x}_\perp-\Vec{x}'_\perp)L_{n-1}^1(\xi^2).
\end{split}
\end{equation}
\end{widetext}

The above result is written as a sum over the Landau levels of the propagator. However, the sum can also be evaluated analytically, with the help of the known integral
\begin{equation}
    \frac{\iu}{x+\iu \epsilon} = \int_0^\infty \dd{s} \e^{\iu s(x + \iu\epsilon)}.
\end{equation}
The $\iu\epsilon$ term makes sure that the integral is convergent and has the correct analytical structure in the complex plane. It will be dropped for the rest of this calculation. Then, the propagator reads
\begin{equation}
\begin{split}
    &S_\mathrm{F}(E,p^z;\Vec{x}_\perp,\Vec{x}'_\perp)=\frac{\e^{\iu\Phi}\e^{-\frac{\xi^2}{2}}}{2\pi\lb}\\
    &\cross\sum_{n=0}^\infty F_n \int_0^\infty \dd{s} \e^{\iu s \left(E^2-(p^z)^2-m_e^2-2n\abs{qB}\right)},
\end{split}
\end{equation}
which can be further simplified with the known sum \citep{gradshtein_table_2015}
\begin{equation}
    \sum_{n=0}^\infty L_n^\alpha(x) z^n = (1-z)^{-(1+\alpha)}\e^{\frac{xz}{z-1}}.
\end{equation}
In our case, $z=\e^{-\iu s 2\abs{qB}}$, and thus, the propagator reads
\begin{equation}
    \begin{split}
        &S_\mathrm{F}(E,p^z;\Vec{x}_\perp,\Vec{x}'_\perp)=\frac{\e^{\iu\Phi}\e^{-\frac{\xi^2}{2}}}{2\pi\lb} \sum_{n=0}^\infty \int_0^\infty \dd{s} \e^{\iu s \left(E^2-(p^z)^2-m_e^2\right)}\\
        &\cross\Big[ \left(\g{0}E-\g{3}p^z+m_e\right)\left( z\mathcal{P}_{-} + \mathcal{P}_{+} \right)\frac{1}{1-z}\\
        &- \frac{\iu}{\lb} \Vec{\gamma}_\perp\vdot(\Vec{x}_\perp-\Vec{x}'_\perp) \frac{z}{(1-z)^2} \Big] \e^{\frac{\xi^2 z}{z-1}}.
    \end{split}
\end{equation}

The above expression can be further simplified by noting that
\begin{equation}
\begin{split}
    \frac{z}{z-1}&=\frac{\e^{\iu x}}{\e^{\iu x}-1} = \frac{1}{2}\left(1-\iu\cot(\frac{x}{2}) \right), \\
    \frac{z}{(z-1)^2} &= -\frac{1}{4}\left(1+\cot[2](\frac{x}{2}), \right)
\end{split}
\end{equation}
where $x=-2s\abs{qB}$. Finally, combining everything, the electron propagator in an external magnetic field reads
\begin{equation}
    \begin{split}
        S_\mathrm{F}(E,p^z&;\Vec{x}_\perp,\Vec{x}'_\perp)=\frac{\e^{\iu\Phi}}{2\pi\lb} \int_0^\infty \dd{s} \e^{\iu s \left(E^2-(p^z)^2-m_e^2\right)}\e^{\iu\frac{\xi^2}{2}\cot(s\abs{qB})}\\
        \cross\Big[& \left(\g{0}E-\g{3}p^z+m_e\right)\frac{\iu}{2}\left( s_\pm\g{1}\g{2} - \cot(s\abs{qB}) \right) \\
        &+ \frac{\iu}{4\lb} \Vec{\gamma}_\perp\vdot(\Vec{x}_\perp-\Vec{x}'_\perp) \left( 1+\cot[2](s\abs{qB}) \right) \Big].
    \end{split}
\end{equation}

\section{\label{app:SynchrotronApp}Scattering amplitude of synchrotron radiation}

As shown in Sec. \ref{sec:Synchrotron}, the scattering amplitude of synchrotron radiation reads
\begin{equation}\label{eq:SynchAmp}
\begin{split}
    \iu T_\mathrm{SR} &= -\iu e\int\dd[4]{x}\ubs{n'}{a'}{\sigma'}{x} \slashed{\varepsilon}_k^*\us{n}{a}{\sigma}{x} \e^{\iu\left(p'+k-p\right)\vdot x} \\
    &= -\iu e \delta^{(3)}(E,p^z,p^y) \int\dd{x} \ubs{n'}{a'}{\sigma'}{x} \slashed{\varepsilon}_k^*\us{n}{a}{\sigma}{x} \e^{-\iu k^x x}. \\
\end{split}
\end{equation}

The above expression consists of a sum of integrals that all have the general form
\begin{equation}
    I_1 = \int\dd{x} H_{\tilde{n}'}\left(\frac{x+\alpha_1}{\lambda_B}\right)H_{\tilde{n}}\left(\frac{x+\alpha_2}{\lambda_B}\right) \e^{-\frac{\left(x+\alpha_1\right)^2}{2\lb}} \e^{-\frac{\left(x+\alpha_2\right)^2}{2\lb}} \e^{\iu \alpha x},
\end{equation}
where
\begin{equation}
    \alpha_1 = \lb (p')^y,\quad \alpha_2 = \lb p^y,\quad \alpha = -k^x,
\end{equation}
$\tilde{n}\in \{n, n-1\}$, and $\tilde{n}'\in \{n', n'-1\}$. Integrals of this form have been evaluated before in, e.g., \citep{mushtukov_compton_2016}. We make the change of variables
\begin{equation}
    u = \frac{x+\frac{\alpha_1+\alpha_2-\iu \alpha \lb}{2}}{\lambda_B},\quad \beta = \frac{\alpha_1 - \alpha_2 -\iu\alpha\lb}{\lambda_B}
\end{equation}
and obtain
\begin{equation}
    I_1 = \lambda_B \e^{-\frac{\abs{\beta}^2}{4}-\iu\frac{\alpha}{2}(\alpha_1+\alpha_2)}\int\dd{u} H_{\tilde{n}'}\left(u+\frac{\beta^*}{2}\right)H_{\tilde{n}}\left(u-\frac{\beta}{2}\right) \e^{-u^2}.
\end{equation}
To proceed, we use Eq. \eqref{eq:HermiteInt} and obtain
\begin{equation}
    I_1 = \begin{cases}
        \lambda_B \e^{-\frac{\abs{\beta}^2}{4}-\iu\frac{\alpha}{2}(\alpha_1+\alpha_2)} 2^{\tilde{n}'}\sqrt{\pi}\tilde{n}!\left(\frac{\beta^*}{2}\right)^{\tilde{n}'-\tilde{n}} L_{\tilde{n}}^{\tilde{n}'-\tilde{n}}\left(\frac{\abs{\beta}^2}{2}\right), & \tilde{n}'\geq \tilde{n} \\
        \lambda_B \e^{-\frac{\abs{\beta}^2}{4}-\iu\frac{\alpha}{2}(\alpha_1+\alpha_2)} 2^{\tilde{n}}\sqrt{\pi}\tilde{n}'!\left(\frac{-\beta}{2}\right)^{\tilde{n}-\tilde{n}'} L_{\tilde{n}'}^{\tilde{n}-\tilde{n}'}\left(\frac{\abs{\beta}^2}{2}\right), & \tilde{n} > \tilde{n}'.
    \end{cases}
\end{equation}
We make the following observations:
\begin{equation}
\begin{split}
    \beta = \iu\lambda_B k_\perp\e^{\iu\phi}&,\quad \abs{\beta}^2=\lb\left((k^x)^2+(k^y)^2\right)=\lb k_\perp^2, \\
    -\iu\frac{\alpha}{2}(\alpha_1+\alpha_2) &= \frac{\iu\lambda_B k^x}{2}\left(p^y+(p')^y\right).
\end{split}
\end{equation}
The result simplifies to
\begin{equation}
    I_1 = \begin{cases}
        \lambda_B 2^\frac{\tilde{n}+\tilde{n}'}{2}\sqrt{\pi \tilde{n}! \tilde{n}'!}\iu^{\tilde{n}'-\tilde{n}}(-1)^{\tilde{n}'}I_{\tilde{n},\tilde{n}'}\left(\frac{\lb k_\perp^2}{2}\right)\e^{\iu(\tilde{n}-\tilde{n}')\phi}, & \tilde{n}' \geq \tilde{n} \\
        \lambda_B 2^\frac{\tilde{n}+\tilde{n}'}{2}\sqrt{\pi \tilde{n}! \tilde{n}'!}\iu^{\tilde{n}'-\tilde{n}}(-1)^{\tilde{n}'}I_{\tilde{n}',\tilde{n}}\left(\frac{\lb k_\perp^2}{2}\right)\e^{\iu(\tilde{n}-\tilde{n}')\phi}, & \tilde{n} > \tilde{n}',
    \end{cases}
\end{equation}
where
\begin{equation}\label{eq:STSpecialFunc}
    I_{n,n'}\left(x\right) = (-1)^n\sqrt{\frac{n!}{n'!}}\e^{-\frac{x}{2}}x^\frac{n'-n}{2}L_n^{n'-n}(x),
\end{equation}
and we have dropped a common phase factor since in the end we are only interested in the squared norm of the amplitude. The total scattering amplitude can be obtained as a sum of these integrals by evaluating the matrix products in Eq. \eqref{eq:SynchAmp}.

From symmetry arguments we know that the square of the matrix element should be independent of $\phi$. Indeed this is the case, since if we take into account the $\phi$ dependence of the polarization vector we find that all terms in the scattering amplitude have the same $\phi$ dependence. This common phase will vanish when taking the squared norm. The final result obtained from this calculation matches the result found in Ref.~\cite{herold_cyclotron_1982} after summing over polarization and spin states of the outgoing photon and electron, respectively.

\section{\label{app:DampingApp}Calculating the decay rate}

In this appendix, I outline the main steps required for taking the zero-temperature limit of the decay rate calculated from the fermion self-energy in Ref.~\citep{ghosh_fermion_2024}. The Feynman diagram of the 1-loop fermion self-energy is shown in Fig.~\ref{fig:Damping}. The imaginary part of the fermion self-energy reads
\begin{widetext}
\begin{equation}\label{eq:SelfEnergy}
    \begin{split}
        &\Im{\Bar{\Sigma}(p_\parallel, \Vec{p}_\perp)} = -4\pi\alpha \sum_{n'=0}^\infty \sum_{s'=\pm}(-1)^{n'}\int\frac{\dd[2]\Vec{k}_\perp}{(2\pi)^2}\e^{-k_\perp^2\lb}\frac{s}{\sqrt{\left(q_\perp^2-(q_\perp^-)^2\right)\left(q_\perp^2-(q_\perp^+)^2\right)}} \\
        &\cross\Big\{\left(sE_{n',k_z^{s'}}\g{0}-(k^z)^{s'}\g{3}\right)\left[\mathcal{P}_- L_{n'}(2k_\perp^2\lb)-\mathcal{P}_+L_{n'-1}(2k_\perp^2\lb)\right] - m_e\left[L_{n'}(2k_\perp^2\lb)-L_{n'-1}(2k_\perp^2\lb)\right] + 2(\Vec{k}_\perp\vdot\Vec{\gamma}_\perp)L_{n'-1}^1(2k_\perp^2\lb) \Big\},
    \end{split}
\end{equation}
where
\begin{equation}
\begin{split}
    (k^z)^{\pm}&=\frac{p^z}{2}\left(1+\frac{2n'\abs{qB}+m^2-q_\perp^2}{p_0^2-p_z^2}\pm\frac{p_0}{p^z(p_0^2-p_z^2)}\sqrt{\left(q_\perp^2-(q_\perp^-)^2\right)\left(q_\perp^2-(q_\perp^+)^2\right)}\right), \\
    E_{n',k_z^\pm} &= \frac{s p_0}{2}\left(1+\frac{2n'\abs{qB}+m^2-q_\perp^2}{p_0^2-p_z^2}\pm\frac{p^z}{p_0(p_0^2-p_z^2)}\sqrt{\left(q_\perp^2-(q_\perp^-)^2\right)\left(q_\perp^2-(q_\perp^+)^2\right)}\right), \\
    q_\perp^\pm &= \abs{\sqrt{m^2+2n'\abs{qB}}\pm\sqrt{p_0^2-p_z^2}}.
\end{split}
\end{equation}
The self-energy can also be obtained by considering the full fermion propagator at 1-loop order. It has the general form
\begin{equation}
    \begin{split}
        &G(x,x')=\iu\left[\iu\slashed{D} - m_e - \Sigma\right]^{-1}\delta^{(4)}(x-x') \\
        &=\iu\Big[(1+\delta v_\parallel)\left(\iu D_0\g{0}-\iu D_3\g{3}\right) -(1+\delta v_\perp)\left(\iu D_1\g{1}+\iu D_2\g{2}\right)+\iu\g{1}\g{2}\left(\Tilde{v}\left(\iu D_0\g{0}-\iu D_3\g{3}\right)-\Tilde{m}_e\right) - (m_e+\delta m_e)\Big]^{-1}\delta^{(4)}(x-x'),
    \end{split}
\end{equation}
where $\delta v_\parallel$, $\delta v_\perp$, $\delta m_e$, $\Tilde{v}$, and $\Tilde{m}_e$ are operators that give corrections to the free fermion propagator $S_\mathrm{F}(x,x')$. In the full fermion propagator, all possible corrections that are allowed by the symmetries of the theory have been added. When the corrections are projected onto the Landau levels, each Landau level receives a unique correction to its properties, determined by $\delta v_{\parallel,n}$, $\delta v_{\perp,n}$, $\delta m_{e,n}$, $\Tilde{v}_n$, and $\Tilde{m}_{e,n}$. At tree level, i.e.,~when using the free fermion propagator, all the corrections vanish.

Using this general form of the full fermion propagator at 1-loop order, the self-energy can be shown to take the form
\begin{equation}
    \begin{split}
        &\Bar{\Sigma}(p_\parallel,\Vec{p}_\perp)=-2\e^{-p_\perp^2\lb}\sum_{n=0}^\infty (-1)^n\left[\delta v_{\parallel,n}(p_\parallel\vdot\gamma_\parallel) + \iu\g{1}\g{2}(p_\parallel\vdot\gamma_\parallel)\Tilde{v}_n-\delta m_{e,n} -\iu\g{1}\g{2}\Tilde{m}_{e,n}\right]\\
        &\cross\left[\mathcal{P}_+L_n(2p_\perp^2\lb) - \mathcal{P}_-L_{n-1}(2p_\perp^2\lb) \right] -4\e^{-p_\perp^2\lb}\sum_{n=0}^\infty (-1)^n\delta v_{\perp,n}(\Vec{\gamma}_\perp\vdot\Vec{p}_\perp)L_{n-1}^1(2p_\perp^2\lb).
    \end{split}
\end{equation}
The values of the correction terms are obtained using the orthogonality of the spin projectors and well-known trace identities of the $\gamma$-matrices:
\begin{equation}
    \begin{split}
        \delta v_{\parallel,n}^+ &= \delta v_{\parallel,n} + s_\pm \Tilde{v}_n = \frac{(-1)^{n+1}\lb}{2\pi p_\parallel^2}\int\dd[2]{\Vec{p}_\perp}\e^{-p_\perp^2\lb}\Tr((p_\parallel\vdot\gamma_\parallel)\mathcal{P}_+\Bar{\Sigma}(p_\parallel,\Vec{p}_\perp))L_n(2p_\perp^2\lb), \\
        \delta v_{\parallel,n}^- &= \delta v_{\parallel,n} - s_\pm \Tilde{v}_n = \frac{(-1)^n\lb}{2\pi p_\parallel^2}\int\dd[2]{\Vec{p}_\perp}\e^{-p_\perp^2\lb}\Tr((p_\parallel\vdot\gamma_\parallel)\mathcal{P}_-\Bar{\Sigma}(p_\parallel,\Vec{p}_\perp))L_{n-1}(2p_\perp^2\lb), \\
        \delta m_{e,n}^+ &= \delta m_{e,n} + s_\pm \Tilde{m}_{e,n} = \frac{(-1)^n\lb}{2\pi}\int\dd[2]{\Vec{p}_\perp}\e^{-p_\perp^2\lb}\Tr(\mathcal{P}_+\Bar{\Sigma}(p_\parallel,\Vec{p}_\perp))L_n(2p_\perp^2\lb), \\
        \delta m_{e,n}^- &= \delta m_{e,n} - s_\pm \Tilde{m}_{e,n} = \frac{(-1)^{n+1}\lb}{2\pi}\int\dd[2]{\Vec{p}_\perp}\e^{-p_\perp^2\lb}\Tr(\mathcal{P}_-\Bar{\Sigma}(p_\parallel,\Vec{p}_\perp))L_{n-1}(2p_\perp^2\lb), \\
        \delta v_{\perp,n} &= \frac{(-1)^n\lambda_B^4}{4\pi n}\int\dd[2]{\Vec{p}_\perp}\e^{-p_\perp^2\lb}\Tr((\Vec{p}_\perp\vdot\Vec{\gamma}_\parallel)\Bar{\Sigma}(p_\parallel,\Vec{p}_\perp))L_{n-1}^1(2p_\perp^2\lb).
    \end{split}
\end{equation}

Combining the above result with the self-energy in Eq.~\eqref{eq:SelfEnergy} yields
\begin{equation}
    \begin{split}
        &\Im{\delta v_{\parallel,n}^+} =\frac{\alpha}{p_\parallel^2}\sum_{n'} \sum_{s'=\pm}\int q_\perp\dd{q_\perp}\Ia{n}{n'-1}{\frac{q_\perp^2\lb}{2}}\frac{\frac{1}{2}\left(p_\parallel^2+2n'\abs{qB}+m_e^2-q_\perp^2\right)}{\sqrt{\left(q_\perp^2-(q_\perp^-)^2\right)\left(q_\perp^2-(q_\perp^+)^2\right)}}, \\
        &\Im{\delta v_{\parallel,n}^-} =\frac{\alpha}{p_\parallel^2}\sum_{n'} \sum_{s'=\pm}\int q_\perp\dd{q_\perp}\Ia{n-1}{n'}{\frac{q_\perp^2\lb}{2}}\frac{ \frac{1}{2}\left(p_\parallel^2+2n'\abs{qB}+m_e^2-q_\perp^2\right)}{\sqrt{\left(q_\perp^2-(q_\perp^-)^2\right)\left(q_\perp^2-(q_\perp^+)^2\right)}}, \\
        &\Im{\delta m_{e,n}^+} =\alpha m_e\sum_{n'} \sum_{s'=\pm} \int q_\perp\dd{q_\perp}\frac{\Ia{n}{n'}{\frac{q_\perp^2\lb}{2}}+\Ia{n}{n'-1}{\frac{q_\perp^2\lb}{2}}}{\sqrt{\left(q_\perp^2-(q_\perp^-)^2\right)\left(q_\perp^2-(q_\perp^+)^2\right)}}, \\
        &\Im{\delta m_{e,n}^-} =\alpha m_e\sum_{n'} \sum_{s'=\pm} \int q_\perp\dd{q_\perp}\frac{\Ia{n-1}{n'}{\frac{q_\perp^2\lb}{2}}+\Ia{n-1}{n'-1}{\frac{q_\perp^2\lb}{2}}}{\sqrt{\left(q_\perp^2-(q_\perp^-)^2\right)\left(q_\perp^2-(q_\perp^+)^2\right)}}, \\
        &\Im{\delta v_{\perp,n}} =\frac{\alpha}{2n}\sum_{n'} \sum_{s'=\pm} \int q_\perp\dd{q_\perp}\frac{\mathcal{I}_2^{n-1,n'-1}\left(\frac{q_\perp^2\lb}{2}\right)}{\sqrt{\left(q_\perp^2-(q_\perp^-)^2\right)\left(q_\perp^2-(q_\perp^+)^2\right)}},
    \end{split}
\end{equation}
where
\begin{equation}
    \mathcal{I}^{n,n'}(x) = I_{n,n'}(x)^2,\quad \mathcal{I}_2^{n,n'}\left(x\right) = \frac{n+n'+2}{2}\left[\Ia{n}{n'}{x}+\Ia{n+1}{n'+1}{x}\right] - \frac{x}{2}\left[\Ia{n+1}{n'}{x} + \Ia{n}{n'+1}{x}\right],
\end{equation}
and the following relation has been used,
\begin{equation}
    sE_{n',k_z^{s'}}p_0-(k^z)^{s'}p^z = \frac{1}{2}\left(p_\parallel^2+2n'\abs{qB}+m_e^2-q_\perp^2\right).
\end{equation}
The integrals are all evaluated over the interval $0 \leq q_\perp \leq q_\perp^-$.

Furthermore, for the numerical evaluation of these integrals, it is useful to make the change of variables
\begin{equation}
    \kappa = \frac{q_\perp^2\lb}{2}.
\end{equation}
With this change of variables, e.g.,~the imaginary part of $\delta v_{\parallel,n}^+$ is
\begin{equation}
    \Im{\delta v_{\parallel,n}^+} =\frac{\alpha}{4p_\parallel^2\lb}\sum_{n'} \sum_{s'=\pm}\int\dd{\kappa}\Ia{n}{n'-1}{\kappa}\frac{\lb\left(p_\parallel^2+m_e^2\right)+2n'-2\kappa}{\sqrt{\left(\kappa-\kappa^-\right)\left(\kappa-\kappa^+\right)}},
\end{equation}
where $\kappa^\pm = \flatfrac{(q_\perp^\pm)^2\lb}{2}$. The other integrals can be evaluated in the same way. Noteworthily, the results are independent of $s'$ and thus the sums over $s'$ can be replaced by $\sum_{s'=\pm}\to 2$.

The decay rates are calculated from the imaginary part of the poles of the full fermion propagator. The position of the poles can be solved from
\begin{equation}
    \left[\left(v_{\parallel,n}^2-\Tilde{v}_n^2\right)^2p_\parallel^2-2nv_{\perp,n}^2\abs{qB}-(m_e+\delta m_{e,n})^2+\Tilde{m}_{e,n}^2\right]^2-4p_\parallel^2\left((m_e+\delta m_{e,n})\Tilde{v}_n-\Tilde{m}_{e,n} v_{\parallel,n}\right)^2=0.
\end{equation}
Assuming the corrections to the propagator are small, the decay rates are
\begin{equation}
    \begin{split}
        \Gamma^\pm_n &\simeq \frac{2\Big(\left(2n\abs{qB}+m_e^2\right)\Im{\delta v_{\parallel,n}} -m_e\Im{\delta m_{e,n}} - 2n\abs{qB}\Im{\delta v_{\perp,n}}\mp s_\pm\sqrt{2n\abs{qB}+m_e^2}\left(m\Im{\Tilde{v}_n}-\Im{\Tilde{m}_{e,n}}\right) \Big)}{\sqrt{2n\abs{qB}+m_e^2+(p^z)^2}}.
    \end{split}
\end{equation}
\end{widetext}
The above relation differs by a factor of 2 from the one given in Ref.~\citep{ghosh_fermion_2024} due to different normalization conventions for the decay rate.

\section{\label{app:ComptonApp}Scattering amplitude calculation of Compton scattering}

This appendix consists of three subsections. In the first subsection, I outline the main steps of the calculation of the cross section of Compton scattering in the LLLa. In the second subsection, I demonstrate how the propagator in the Schwinger proper time formalism can be used to calculate the Compton scattering cross section when the external electron legs are restricted to the LLL. Finally, in the third subsection I give the final results of the scattering amplitudes of Compton scattering in the general case, where there are no restrictions on the Landau levels of real and virtual fermions.

\subsection{Lowest Landau level approximation} \label{app:LLL}
In the LLLa, it is assumed that the electron propagator as well as the external electron legs are on the LLL. This is the same approximation that is used in, e.g., Ref.~\citep{kostenko_qed_2018}. In practice, the LLLa means using only the $n=0$ term of the sum in Eq.~\eqref{eq:PropLevels}. Then, the propagator reads
\begin{equation}
    S_{\mathrm{F},0}(E_I,p^z_I; \Vec{x}'_\perp, \Vec{x}_\perp)=\iu \frac{\e^{\iu\Phi(x',x)}\e^{-\frac{\xi^2}{2}}}{2\pi\lb} \frac{\left(\g{0}E_I-\g{3}p_I^z+m_e\right)}{E_I^2-(p_I^z)^2-m_e^2} \mathcal{P}_+.
\end{equation}

Making use of the Feynman rules defined in Sec.~\ref{sec:FeynRules}, the scattering amplitude for the $s$-channel diagram reads
\begin{equation}
\begin{split}
    \iu T_s &= (-\iu e)^2\int \dd[4]{x'}\dd[4]{x}\ubs{0}{a'}{-1}{x'}\e^{\iu p'\vdot x'}\slashed{\varepsilon}^*\e^{\iu k'\vdot x'}\\
    &\cross S_{\mathrm{F},0}(x',x)\slashed{\varepsilon}\e^{-\iu k\vdot x}\us{0}{a}{-1}{x}\e^{-\iu p\vdot x},
\end{split}
\end{equation}
where $S_{\mathrm{F},0}(x',x)$ is the position space electron propagator. It is most convenient to work in a mixed representation, with the $t$- and $z$-components of the propagator given in momentum space,
\begin{equation}
    S_{\mathrm{F},0}(x',x)=\int\frac{\dd{E_I}\dd{p^z_I}}{(2\pi)^2}S_{\mathrm{F},0}(E_I,p^z_I;\Vec{x}'_\perp,\Vec{x}_\perp)\e^{-\iu(E_I (t-t') - p^z_I (z-z'))}.
\end{equation}
The calculation of the integrals over $t$, $t'$, and $E_I$ yields a Dirac $\delta$-function that enforces conservation of energy,
\begin{equation}
    \begin{split}
        \int&\frac{\dd{E_I}}{2\pi} \int \dd{t}\e^{-\iu t\left(E + \omega -E_I\right)}\int\dd{t'}\e^{\iu t'\left(E' + \omega' -E_I\right)} \\
        =& \; 2\pi \delta(E + \omega - E' - \omega').
    \end{split}
\end{equation}
Similarly, the $z$-part enforces conservation of momentum in the $z$-direction. Therefore, the $s$-channel scattering amplitude reads
\begin{equation}
\begin{split}
    \iu T_s &= -\frac{2\pi\iu e^2}{\pi^\frac{1}{2}\lambda_B^3}\delta(E + \omega - E' - \omega')\delta(p^z+k^z-(p')^z-(k')^z) \\
    &\cross\int \dd[2]{x_\perp}\dd[2]{x_\perp'}\e^{\iu (p^y+k^y)y}\e^{-\iu \left((p')^y+(k')^y\right)y'}\e^{\iu k^x x}\e^{-\iu (k')^x x'}\\
    &\cross\e^{-\frac{(x-a)^2}{2\lb}}\e^{-\frac{(x'-a')^2}{2\lb}} \e^{-\iu s_\pm\frac{(x+x')(y-y')}{2\lb}}\e^{-\frac{(x-x')^2+(y-y')^2}{4\lb}} \\
    &\cross \left(C^{\mathrm{ST'}}\right)^\dagger\g{0}\slashed{\varepsilon}_{k'}^*\frac{\left(\g{0}E_I-\g{3}p_I^z+m_e\right)}{E_I^2-(p_I^z)^2-m_e^2} \mathcal{P}_+\slashed{\varepsilon}_kC^{\mathrm{ST}}.
\end{split}
\end{equation}
This integral is most naturally calculated by performing a change of variables into center-of-mass (CoM) coordinates, defined as:
\begin{equation}
    \begin{split}
        y_r &= y-y',\quad Y = \frac{y+y'}{2}, \\
        x_r &= x-x',\quad X = \frac{x+x'}{2}.
    \end{split}
\end{equation}
The Jacobian of the transformation is unity. In CoM coordinates, the $Y$-dependent part of the scattering amplitude reads
\begin{equation}
    \int \dd{Y}\e^{\iu (p^y+k^y-(p')^y-(k')^y)Y}=2\pi\delta(p^y+k^y-(p')^y-(k')^y),
\end{equation}
which enforces momentum conservation in the $y$-direction. 

The three remaining integrals are all Gaussian and we obtain the result for the $s$-channel scattering amplitude
\begin{equation}
    \begin{split}
        \iu T_s &=-(2\pi)^3\iu e^2\delta^{(3)}(E,p^z,p^y) \e^{-\frac{1}{4}\lb\left(k_\perp^2 + (k')_\perp^2\right)} \e^{\iu \Theta} \e^{\iu \alpha} \\
        &\cross \left(C^{\mathrm{ST'}}\right)^\dagger\g{0}\slashed{\varepsilon}_{k'}^*\frac{\left(\g{0}E_I-\g{3}p_I^z+m_e\right)}{E_I^2-(p_I^z)^2-m_e^2} \mathcal{P}_+\slashed{\varepsilon}_kC^{\mathrm{ST}},
    \end{split}
\end{equation}
where $s_\pm=-1$ for the electron,
\begin{equation}\label{eq:ThetaAlpha}
    \begin{split}
       k_\perp^2 &= (k^x)^2 + (k^y)^2 = \omega^2\sin[2](\theta), \\
       \Theta &= \frac{1}{2}\lb k_\perp (k')_\perp\sin(\phi-\phi'), \\
       \alpha &= \frac{1}{2} \lb \left(k_x'-k_x\right)\left(p_y+p_y'\right).
    \end{split}
\end{equation}

The $u$-channel scattering amplitude can be obtained using crossing symmetry \citep{peskin_introduction_1995}. The two amplitudes are identical apart from the substitutions $k\leftrightarrow-k'$ and $\slashed{\varepsilon}\leftrightarrow\slashed{\varepsilon}^*$, yielding 
\begin{equation}
\begin{split}
    \iu T_u &= -(2\pi)^3\iu e^2\delta^{(3)}(E,p^z,p^y) \e^{-\frac{1}{4}\lb\left(k_\perp^2 + (k')_\perp^2\right)} \e^{-\iu \Theta} \e^{\iu \alpha} \\
    &\cross \left(C^{\mathrm{ST'}}\right)^\dagger\g{0}\slashed{\varepsilon}_k\frac{\left(\g{0}E_I-\g{3}p_I^z+m_e\right)}{E_I^2-(p_I^z)^2-m_e^2} \mathcal{P}_+\slashed{\varepsilon}_{k'}^*C^{\mathrm{ST}}.
\end{split}
\end{equation}
Remarkably, the only differences between the $s$ and $u$-channels can be seen in the sign of the $\Theta$-phase, the order of the polarization vectors, and the 4-momentum in the propagator. Both channels have the common phase factor $\e^{\iu\alpha}$. Since the physical quantity of interest is the squared norm of the matrix element, the common phase factor can be dropped.

Finally, all that is left is to calculate the part of the amplitudes that includes the Dirac and Lorentz structure. In the $s$-channel $E_I=E+\omega$ and $p^z_I=p^z+k^z$ and we obtain in the ERF
\begin{equation}
\begin{split}
    &\left(C^{\mathrm{ST'}}\right)^\dagger\g{0}\slashed{\varepsilon}_{k'}^*\frac{\left(\g{0}E_I-\g{3}p_I^z+m_e\right)}{E_I^2-(p_I^z)^2-m_e^2} \mathcal{P}_+\slashed{\varepsilon}_kC^{\mathrm{ST}} \\
    &=\frac{\sqrt{2m_e}\varepsilon_k^z (\varepsilon_{k'}^z)^*}{\sqrt{E'+m_e}}\frac{\left(\omega(E'+m_e)+k^z (p')^z\right)}{(\omega+m_e)^2-(k^z)^2-m_e^2}\\
    &\equiv  \frac{\sqrt{2m_e}\varepsilon_k^z (\varepsilon_{k'}^z)^*}{\sqrt{E'+m_e}} F_+^{(0)}.
\end{split}
\end{equation}
Similarly, in the $u$-channel $E_I=E-\omega'$ and $p^z_I=p^z-(k')^z$ and thus 
\begin{equation}
\begin{split}   
    &\left(C^{\mathrm{ST'}}\right)^\dagger\g{0}\slashed{\varepsilon}_k\frac{\left(\g{0}E_I-\g{3}p_I^z+m_e\right)}{E_I^2-(p_I^z)^2-m_e^2} \mathcal{P}_+\slashed{\varepsilon}_{k'}^*C^{\mathrm{ST}} \\
    &=\frac{\sqrt{2m_e}\varepsilon_k^z (\varepsilon_{k'}^z)^*}{\sqrt{E'+m_e}}\frac{\left(-\omega'(E'+m_e)-(k')^z (p')^z\right)}{(m_e-\omega')^2-(k_z')^2-m_e^2}\\
    &\equiv \frac{\sqrt{2m_e}\varepsilon_k^z (\varepsilon_{k'}^z)^*}{\sqrt{E'+m_e}} F_-^{(0)}.
\end{split}
\end{equation}
These scattering amplitudes were also obtained in Ref.~\citep[eqs.~37--38]{kostenko_qed_2018} except that they have fixed their gauge such that $k^x=0$.

The differential cross section is obtained from the squared norm of the matrix element, as outlined in Sec.~\ref{sec:CrossSec}. The only dependence in the obtained result on the angles $\phi$ and $\phi'$ is in the variable $\Theta$, that depends only on the difference $\Delta\phi=\phi'-\phi$. Thus, integrating over $\phi'$ is equivalent to integrating over $\Delta\phi$. The integral yields a Bessel function,
\begin{equation} \label{eq:BesselInt}
\begin{split}
    &\int_0^{2\pi} \dd{\phi}\e^{\iu(n\phi\pm x\sin{\phi})}=(\mp 1)^n 2\pi J_n(x),\\
    &J_{-n}(x)=(-1)^nJ_n(x),\quad n\in\mathbb{Z},
\end{split}
\end{equation}
and we obtain the differential cross section
\begin{equation}
\begin{split}
    &\dv{\sigma}{\cos{\theta'}}=\frac{3\sigma_\mathrm{T}}{8}\frac{\omega'}{\omega}\frac{m_e^2\abs{\varepsilon_k^z}^2\abs{\varepsilon_{k'}^z}^2}{(E'+m_e)(E'-(p')^z\cos{\theta'})} \\
    &\cross \e^{-\frac{1}{2}\lb\left(k_\perp^2 + (k')_\perp^2\right)}\left(F_+^2+F_-^2+2 F_+ F_- J_0(\lb k_\perp k_\perp')\right),
\end{split}
\end{equation}
where $\sigma_\mathrm{T}=\alpha_e^2\flatfrac{8\pi}{(3m_e^2)}=r_e^2\flatfrac{8\pi}{3}$ is the Thomson cross section and $r_e=\flatfrac{\alpha_e}{m_e}$ is the classical electron radius. Finally, the total cross section can be obtained by integrating the above expression numerically. Since the differential cross section is proportional to the $z$-component of the polarization vector, the differential cross section vanishes if either photon is in the X-mode.

\subsection{\label{app:ComptonSchwinger}Cross section with the Schwinger proper time propagator}
In this section, the fermion propagator in the Schwinger-proper-time form is used to calculate the scattering amplitudes of Compton scattering. The calculation follows the same procedure introduced in Sec.~\ref{sec:CrossSec} and the integrals over $z$, $z'$, $t$, $t'$, and $Y$ are calculated in the same way as done in Sec.~\ref{app:LLL}. The remaining $y_r$, $X$, and $x_r$ integrals can all be written as linear combinations of the following integrals (Gaussian and Fresnel integrals):
\begin{equation}
\begin{split}
    I_1(A,B,C)&=\int_{-\infty}^\infty \dd{x} \e^{-Ax^2+Bx+C}=\sqrt{\frac{\pi}{A}}\e^{\frac{B^2}{4A}+C}, \\
    I_2(A,B,C)&=\int_{-\infty}^\infty \dd{x} x\e^{-Ax^2+Bx+C}=\frac{B}{2A}\sqrt{\frac{\pi}{A}}\e^{\frac{B^2}{4A}+C}, \\
    I_3(A,B,C)&=\int_{-\infty}^\infty \dd{x} \e^{\iu Ax^2+\iu Bx+C}=\sqrt{\frac{\pi}{\abs{A}}}\e^{\frac{-\iu B^2}{4A}+C+\sign(A)\iu\frac{\pi}{4}}, \\
    I_4(A,B,C)&=\int_{-\infty}^\infty \dd{x} x\e^{\iu Ax^2+\iu Bx+C}\hspace{-0.25em}=\hspace{-0.25em}-\frac{B}{2A}\sqrt{\frac{\pi}{\abs{A}}}\e^{\frac{-\iu B^2}{4A}+C+\sign(A)\iu\frac{\pi}{4}}.
\end{split}
\end{equation}
\vspace{2em}
For $I_1$ and $I_2$ we require that $\Re(A)>0$ and for $I_3$ and $I_4$ we require $A,B\in \mathbb{R}$. Note, that $I_4$ is not a convergent integral and should be thought of as the limiting case of $I_2$, when $\Re(A),\Re(B)\to 0^+$.

The result reads
\begin{equation}
    \begin{split}
        &\iu T_s = -N e^2 \delta^3(E,p^y,p^z)\frac{(2\pi)^2}{\lb} \int_0^\infty\dd{s}\e^{\iu s\left(E_I^2-(p^z_I)^2-m_e^2\right)}\\
        &\cross\frac{\pi}{\sqrt{A_1 A_2}}\e^{\frac{B_1^2}{4 A_1}+\frac{B_2^2}{4 A_2}+C} \Tilde{C}^\dagger\g{0}\slashed{\varepsilon}^*\left(\beta_1 + \beta_2 \frac{B_1}{2 A_1}+\beta_3\frac{B_2}{2 A_2}\right)\slashed{\varepsilon}\Tilde{C},
    \end{split}
\end{equation}
where 
\begin{widetext}
\begin{equation}
\begin{split}
    A_1 &= \frac{1}{\lb}\left(1+\iu\tan(s\abs{eB})\right),\quad B_1 = \iu(k^x-(k')^x)+2\iu s_\pm(p^y+k^y)\tan(s\abs{eB})+\frac{a+a'}{\lb}, \\
    A_2&\equiv\frac{1}{4\lb}\left(1-\iu\cot(s\abs{eB})\right),\quad B_2 \equiv\frac{1}{2}\left(\iu(k^x+(k')^x)+\frac{a-a'}{\lb}\right),\quad C = -\frac{a^2+a'^2}{2\lb}-\iu\lb(p^y+k^y)^2\tan(s\abs{eB}) + \iu s_c\frac{\pi}{4}, \\
    s_c &= \sign(\tan(s\abs{eB})),\quad \beta_1 =\sqrt{4\pi\lb\abs{\tan(s\abs{eB})}}\left(\alpha_1 - 2\lb \alpha_2\tan(s\abs{eB})\left(p^y + k^y\right)\right), \\
    \beta_2 &=2s_\pm \alpha_2 \sqrt{4\pi\lb\abs{\tan(s\abs{eB})}}\tan(s\abs{eB}),\quad \beta_3=\alpha_3\sqrt{4\pi\lb\abs{\tan(s\abs{eB})}}, \\
    \alpha_1 &= \left(\g{0}E_I-\g{3}p^z_I+m_e\right) \frac{\iu}{2} \left( s_\pm\g{1}\g{2}-\cot(s\abs{eB}) \right),\quad \alpha_2 =-\frac{\iu}{4\lb}\g{2}\left(1+\cot[2](s\abs{eB})\right),\quad
    \alpha_3 =-\frac{\iu}{4\lb}\g{1}\left(1+\cot[2](s\abs{eB})\right),
\end{split}    
\end{equation}
and $N$ includes the normalization constants of the wave functions. The amplitude can also be written with the Dirac structure more transparent,
\begin{equation}\label{eq:SchwingerZeta}
    \begin{split}
        \iu T_s &= -N e^2 \delta^3(E,p^z,p^y)\frac{(2\pi)^2}{\lb} \int_0^\infty\dd{s}\e^{\iu s\left(E_I^2-(p^z_I)^2-m_e^2\right)}\frac{\pi}{\sqrt{A_1 A_2}}\e^{\frac{B_1^2}{4 A_1}+\frac{B_2^2}{4 A_2}+C}\\
        &\cross\sqrt{4\pi\lb\abs{\tan(s\abs{eB})}}\varepsilon^*_\mu\varepsilon_\nu\Tilde{C}^\dagger\g{0}\g{\mu}\left(\zeta + \zeta_0 \g{0} + \zeta_1 \g{1} + \zeta_2 \g{2} +\zeta_3 \g{3} + \zeta_{12}\g{1}\g{2} + \zeta_{012} \g{0}\g{1}\g{2} + \zeta_{312} \g{3}\g{1}\g{2} \right)\g{\nu}\Tilde{C},
    \end{split}
\end{equation}
where
\begin{equation}
    \begin{split}
        \zeta_0 &= -\iu\frac{E_I}{2}\cot(s\abs{eB}),\quad \zeta_1 = -\iu\frac{B_2\csc[2](s\abs{eB})}{8A_2\lb},\quad \zeta_2 = \iu\tan(s\abs{eB})\csc[2](s\abs{eB})\frac{1}{4\lb}\left(-s_\pm\frac{B_1}{A_1}+2\lb(p^y+k^y)\right) \\
        \zeta_3 &= \iu\frac{p^z_I}{2} \cot(s\abs{eB}), \quad\zeta_{12}=\frac{\iu s_\pm m_e}{2}, \quad \zeta_{012} = \frac{\iu E_I s_\pm}{2},\quad \zeta_{312} = -\frac{\iu p^z_I s_\pm}{2},\quad \zeta = -\iu\frac{m_e}{2}\cot(s\abs{eB}).
    \end{split}
\end{equation}
In the further simplifications, the following trigonometric identities are used:
\begin{equation}
    \begin{split}
        &\iu + \cot{x} = \frac{\iu\tan{x} + 1}{\tan{x}} = \frac{\tan{x}-\iu}{-\iu\tan{x}},\quad \frac{1-\iu\tan{x}}{1+\iu\tan{x}}=\frac{1+\iu\cot{x}}{1-\iu\cot{x}}=\e^{-2\iu x}.
    \end{split}
\end{equation}

The exponent in Eq.~\eqref{eq:SchwingerZeta} reads
\begin{equation} \label{eq:FullExpSimplification}
\begin{split}
    &\frac{B_1^2}{4 A_1}+\frac{B_2^2}{4 A_2}+C = \frac{\iu\lb \left(-k^x+(k')^x-\iu k^y +\iu (k')^y +2k^y \tan{\tau} \right)^2}{4\left(-\iu+\tan{\tau}\right)} \\
    &- \frac{\iu\lb \left(k^x+(k')^x-\iu k^y +\iu (k')^y \right)^2}{4\left(\iu+\cot{\tau}\right)} + \frac{-2\lb\left(k^y-(k')^y\right)^2-4\iu\lb (k^y)^2\tan{\tau}+\iu\pi s_c}{4}\\
    =&-\frac{1}{4}\lb\left(k_\perp^2+(k')_\perp^2\right) +\iu\Theta + \iu\alpha + \iu s_c\frac{\pi}{4} + \frac{1}{2}\lb\e^{-2\iu\tau}\omega\omega' \sin{\theta}\sin{\theta'}\e^{-\iu(\phi-\phi')},
\end{split}
\end{equation}
where $\tau=\flatfrac{s}{\lb}$,  $s_\pm=-1$, 
and $\Theta$ and $\alpha$ are defined as in the LLLa calculation in Sec.~\ref{app:LLL}. Importantly, the parts independent of $\tau$ are the same as when the amplitude is calculated by adding together a finite amount of Landau levels.

Additionally, it can be shown that
\begin{equation}
\begin{split}
    \frac{\pi}{\sqrt{A_1 A_2}}&=\sqrt{\frac{4\pi^2\lambda_B^4}{(1+\iu\tan{\tau})(1-\iu\cot{\tau})}}=\sqrt{\frac{- 16\iu\pi^3\lambda_B^6}{4\pi\lb\tan{\tau}(1-\iu\cot{\tau})^2}}.
\end{split}
\end{equation}
Finally, all that is left is to calculate the Dirac and Lorentz structure of the amplitude. The result reads
\begin{equation}
    \begin{split}
        &\iu T_s = -N e^2 \delta^3(E,p^z,p^y)\frac{1}{2\pi\lb} \int_0^\infty\dd{s}\e^{\iu s\left(E_I^2-(p^z_I)^2-m_e^2\right)}\frac{\pi}{\sqrt{A_1 A_2}}\e^{\frac{B_1^2}{4 A_1}+\frac{B_2^2}{4 A_2}+C} \\
        &\Big[ -\zeta\left(\Tilde{C}_2(\Tilde{C}_2')^*-\Tilde{C}_4(\Tilde{C}_4')^*\right)\left(\varepsilon_k^- \varepsilon_{k'}^{-*} + \varepsilon_k^z \varepsilon_{k'}^{z*}\right) + \zeta_0\left(\Tilde{C}_2(\Tilde{C}_2')^*+\Tilde{C}_4(\Tilde{C}_4')^*\right)\left(\varepsilon_k^- \varepsilon_{k'}^{-*} + \varepsilon_k^z \varepsilon_{k'}^{z*}\right) \\
        &+ \zeta_1\left(\Tilde{C}_4 (\Tilde{C}_2')^*+\Tilde{C}_2 (\Tilde{C}_4')^*\right)\left(\varepsilon_k^z \varepsilon_{k'}^{-*} + \varepsilon_k^- \varepsilon_{k'}^{z*}\right) -\iu\zeta_2\left(\Tilde{C}_4 (\Tilde{C}_2')^*+\Tilde{C}_2 (\Tilde{C}_4')^*\right)\left(\varepsilon_k^z \varepsilon_{k'}^{-*} - \varepsilon_k^- \varepsilon_{k'}^{z*}\right) \\
        &- \zeta_3\left(\Tilde{C}_4 (\Tilde{C}_2')^*+\Tilde{C}_2 (\Tilde{C}_4')^*\right)\left(\varepsilon_k^- \varepsilon_{k'}^{-*} - \varepsilon_k^z \varepsilon_{k'}^{z*}\right) +\iu\zeta_{12}\left(\Tilde{C}_2(\Tilde{C}_2')^*-\Tilde{C}_4(\Tilde{C}_4')^*\right)\left(\varepsilon_k^- \varepsilon_{k'}^{-*} - \varepsilon_k^z \varepsilon_{k'}^{z*}\right) \\
        &-\iu \zeta_{012} \left(\Tilde{C}_2(\Tilde{C}_2')^*+\Tilde{C}_4(\Tilde{C}_4')^*\right) \left(\varepsilon_k^- \varepsilon_{k'}^{-*} - \varepsilon_k^z \varepsilon_{k'}^{z*}\right) \\
        &+\iu \zeta_{312} \left(\Tilde{C}_4 (\Tilde{C}_2')^*+\Tilde{C}_2 (\Tilde{C}_4')^*\right) \left(\varepsilon_k^- \varepsilon_{k'}^{-*} + \varepsilon_k^z \varepsilon_{k'}^{z*}\right) \Big].
    \end{split}
\end{equation}
These polarization vectors can be in either the O-mode or the X-mode. The cross section can be calculated for different combinations of polarizations and can also be averaged over incoming and/or outgoing polarizations. 

For example, the scattering amplitude for both photons in the O-mode reads
\begin{equation}
    \begin{split}
        &\iu T_s^{\mathrm{OO}}= -\frac{(2\pi)^3 e^2\sqrt{2m_e}}{\sqrt{E'+m_e}}\delta^3(E,p^z,p^y)\e^{-\frac{1}{4}\lb\left(k_\perp^2+(k')_\perp^2\right) +\iu\Theta + \iu\alpha}\\
        &\cross\int_0^\infty\dd{s}\e^{\iu s\left(E_T^2-Z_T^2-m_e^2\right)} \exp(\frac{1}{2}\lb\e^{-2\iu \flatfrac{s}{\lb}}\omega\omega' \sin{\theta}\sin{\theta'}\e^{\iu(\phi'-\phi)})\Bigg[ \sin{\theta}\sin{\theta'}\left(E_T^2-E_T\omega' +m_e\omega'-m_e^2+Z_T^2-Z_T\omega'\cos{\theta'}\right) \\
        &- \e^{-\frac{2\iu s}{\lb}} \e^{\iu(\phi'-\phi)}\Big(\left( \omega'\sin[2](\theta')\cos{\theta} + \omega\cos{\theta'}\sin[2](\theta)\right)(\omega'\cos{\theta'}-Z_T)+ \cos{\theta}\cos{\theta'} \left( E_T^2-E_T\omega' +m_e\omega'-m_e^2-Z_T^2 +Z_T\omega'\cos{\theta'} \right)\Big)\Bigg],
    \end{split}
\end{equation}
where $E_T=E+\omega$ and $Z_T=p^z+k^z$.

The scattering amplitude in the $u$-channel is obtained with the usual substitutions and reads
\begin{equation}
    \begin{split}
        &\iu T_u^{\mathrm{OO}}= -\frac{(2\pi)^3e^2\sqrt{2m_e}}{\sqrt{E'+m_e}}\delta^3(E,p^z,p^y)\e^{-\frac{1}{4}\lb\left(k_\perp^2+(k')_\perp^2\right) - \iu\Theta + \iu\alpha} \\
        &\cross\int_0^\infty\dd{s}\e^{\iu s\left((E-\omega')^2-((k')^z)^2-m_e^2\right)} \exp(\frac{1}{2}\lb\e^{-2\iu \flatfrac{s}{\lb}}\omega\omega' \sin{\theta}\sin{\theta'}\e^{-\iu(\phi'-\phi)})\Bigg[ -\omega'\sin{\theta}\sin{\theta'}\left(E_T-\omega' +m_e +Z_T\cos{\theta'} -\omega'\cos[2](\theta')\right) \\
        &+ \e^{-\frac{2\iu s}{\lb}}\e^{-\iu(\phi'-\phi)}\Big(\left( \omega'\sin[2](\theta')\cos{\theta} + \omega\cos{\theta'}\sin[2](\theta)\right)(\omega'\cos{\theta'}-Z_T) + \omega'\cos{\theta}\cos{\theta'} \left( E_T-\omega' +m_e -Z_T\cos{\theta'} +\omega'\cos[2](\theta') \right)\Big)\Bigg].
    \end{split}
\end{equation}
For the case $\theta=0=\theta'$, it is straightforward to show that these scattering amplitudes match with the Landau level projected case.

\subsection{\label{app:ComptonResult}Final scattering amplitudes}
We have defined the following shorthand notation:
\begin{equation}
\begin{split}
    \m{s}{\pm}{1}{1} &= \CC{1}{1}\left(E_I-m_e\right)+\CC{3}{3}\left(E_I+m_e\right)\mp\left(\CC{1}{3}+\CC{3}{1}\right)p_I^z = \m{u}{\pm}{1}{1}, \\
    \m{s}{\pm}{1}{2} &= \mp\CC{2}{1}\left(E_I-m_e\right)\mp\CC{4}{3}\left(E_I+m_e\right)+\left(\CC{2}{3}+\CC{4}{1}\right)p_I^z = \m{u}{\pm}{1}{2}, \\
    \m{s}{\pm}{1}{3} &= \mp\CC{1}{2}\left(E_I-m_e\right)\mp\CC{3}{4}\left(E_I+m_e\right)+\left(\CC{3}{2}+\CC{1}{4}\right)p_I^z = \m{u}{\pm}{1}{3}, \\ 
    \m{s}{\pm}{1}{4} &= \CC{2}{2}\left(E_I-m_e\right)+\CC{4}{4}\left(E_I+m_e\right)\mp\left(\CC{2}{4}+\CC{4}{2}\right)p_I^z = \m{u}{\pm}{1}{4}, \\
    \m{s}{\pm}{2}{1} &= -\left(\CC{3}{1}+\CC{1}{3}\right) = -\m{u}{\pm}{2}{1},\quad \m{s}{\pm}{2}{2} = \mp\left(\CC{2}{3}+\CC{4}{1}\right) = -\m{u}{\pm}{2}{2}, \\
    \m{s}{\pm}{2}{3} &= \pm\left(\CC{3}{2}+\CC{1}{4}\right) = -\m{u}{\pm}{2}{3},\quad \m{s}{\pm}{2}{4} = \left(\CC{4}{2}+\CC{2}{4}\right) = - \m{u}{\pm}{2}{4}.
\end{split}
\end{equation}
The final scattering amplitudes read
\onecolumngrid
\clearpage
\twocolumngrid
\end{widetext}
\small
\begin{equation}\label{eq:TsAmp}
    \begin{split}
        &\iu T_s^{(n)} = -(2\pi)^3\iu e^2\delta^{(3)}(E,p^z,p^y)\frac{\iu^{n_\mathrm{i}-n_\mathrm{f}}\e^{\iu \Theta} \e^{\iu \alpha}}{E_I^2-(p_I^z)^2-m_e^2-2eBn}\Bigg\{ \\
        &\m{s}{+}{1}{1}\f{+}{\theta}\f{+}{\theta'}^*\Ik{n_\mathrm{i}-1}{n}\Ikp{n_\mathrm{f}-1}{n} \\
        &+\iu \m{s}{+}{1}{2}\f{z}{\theta}\f{+}{\theta'}^*\Ik{n_\mathrm{i}}{n}\Ikp{n_\mathrm{f}-1}{n} \\
        &-\iu \m{s}{+}{1}{3}\f{+}{\theta}\f{z}{\theta'}^*\Ik{n_\mathrm{i}-1}{n}\Ikp{n_\mathrm{f}}{n} \\
        &+ \m{s}{+}{1}{4}\f{z}{\theta}\f{z}{\theta'}^*\Ik{n_\mathrm{i}}{n}\Ikp{n_\mathrm{f}}{n}\\
        &+\m{s}{-}{1}{1}\f{z}{\theta}\f{z}{\theta'}^*\Ik{n_\mathrm{i}-1}{n-1}\Ikp{n_\mathrm{f}-1}{n-1} \\
        &+\iu \m{s}{-}{1}{2}\f{-}{\theta}\f{z}{\theta'}^*\Ik{n_\mathrm{i}}{n-1}\Ikp{n_\mathrm{f}-1}{n-1} \\
        &-\iu \m{s}{-}{1}{3}\f{z}{\theta}\f{-}{\theta'}^*\Ik{n_\mathrm{i}-1}{n-1}\Ikp{n_\mathrm{f}}{n-1} \\
        &+ \m{s}{-}{1}{4}\f{-}{\theta}\f{-}{\theta'}^*\Ik{n_\mathrm{i}}{n-1}\Ikp{n_\mathrm{f}}{n-1} \\
        &+\sqrt{2bn}\Bigg[\m{s}{-}{2}{1}\f{+}{\theta}\f{z}{\theta'}^*\Ik{n_\mathrm{i}-1}{n}\Ikp{n_\mathrm{f}-1}{n-1} \\
        &+\iu \m{s}{-}{2}{2}\f{z}{\theta}\f{z}{\theta'}^*\Ik{n_\mathrm{i}}{n}\Ikp{n_\mathrm{f}-1}{n-1} \\
        &-\iu \m{s}{-}{2}{3}\f{+}{\theta}\f{-}{\theta'}^*\Ik{n_\mathrm{i}-1}{n}\Ikp{n_\mathrm{f}}{n-1} \\
        &+ \m{s}{-}{2}{4}\f{z}{\theta}\f{-}{\theta'}^*\Ik{n_\mathrm{i}}{n}\Ikp{n_\mathrm{f}}{n-1} \\
        &+ \m{s}{+}{2}{1}\f{z}{\theta}\f{+}{\theta'}^*\Ik{n_\mathrm{i}-1}{n-1}\Ikp{n_\mathrm{f}-1}{n} \\
        &+\iu \m{s}{+}{2}{2}\f{-}{\theta}\f{+}{\theta'}^*\Ik{n_\mathrm{i}}{n-1}\Ikp{n_\mathrm{f}-1}{n} \\
        &-\iu \m{s}{+}{2}{3}\f{z}{\theta}\f{z}{\theta'}^*\Ik{n_\mathrm{i}-1}{n-1}\Ikp{n_\mathrm{f}}{n} \\
        &+ \m{s}{+}{2}{4}\f{-}{\theta}\f{z}{\theta'}^*\Ik{n_\mathrm{i}}{n-1}\Ikp{n_\mathrm{f}}{n}\Bigg]\Bigg\}\\
    \end{split}
\end{equation}

\begin{equation}\label{eq:TuAmp}
    \begin{split}
        &\iu T_u^{(n)} = -(2\pi)^3\iu e^2\delta^{(3)}(E,p^z,p^y)\frac{\iu^{n_\mathrm{f}-n_\mathrm{i}}\e^{-\iu \Theta} \e^{\iu \alpha}}{E_I^2-(p_I^z)^2-m_e^2-2eBn}\Bigg\{ \\
        &\m{u}{+}{1}{1}\f{-}{\theta}\f{-}{\theta'}^*\Ikp{n_\mathrm{i}-1}{n}\Ik{n_\mathrm{f}-1}{n} \\
        &-\iu \m{u}{+}{1}{2}\f{-}{\theta}\f{z}{\theta'}^*\Ikp{n_\mathrm{i}}{n}\Ik{n_\mathrm{f}-1}{n} \\
        &+\iu \m{u}{+}{1}{3}\f{z}{\theta}\f{-}{\theta'}^*\Ikp{n_\mathrm{i}-1}{n}\Ik{n_\mathrm{f}}{n} \\
        &+ \m{u}{+}{1}{4}\f{z}{\theta}\f{z}{\theta'}^*\Ikp{n_\mathrm{i}}{n}\Ik{n_\mathrm{f}}{n}\\
        &+\m{u}{-}{1}{1}\f{z}{\theta}\f{z}{\theta'}^*\Ikp{n_\mathrm{i}-1}{n-1}\Ik{n_\mathrm{f}-1}{n-1} \\
        &-\iu \m{u}{-}{1}{2}\f{z}{\theta}\f{+}{\theta'}^*\Ikp{n_\mathrm{i}}{n-1}\Ik{n_\mathrm{f}-1}{n-1} \\
        &+\iu \m{u}{-}{1}{3}\f{+}{\theta}\f{z}{\theta'}^*\Ikp{n_\mathrm{i}-1}{n-1}\Ik{n_\mathrm{f}}{n-1} \\
        &+ \m{u}{-}{1}{4}\f{+}{\theta}\f{+}{\theta'}^*\Ikp{n_\mathrm{i}}{n-1}\Ik{n_\mathrm{f}}{n-1} \\
        &+\sqrt{2bn}\Bigg[\m{u}{-}{2}{1}\f{z}{\theta}\f{-}{\theta'}^*\Ikp{n_\mathrm{i}-1}{n}\Ik{n_\mathrm{f}-1}{n-1} \\
        &-\iu \m{u}{-}{2}{2}\f{z}{\theta}\f{z}{\theta'}^*\Ikp{n_\mathrm{i}}{n}\Ik{n_\mathrm{f}-1}{n-1} \\
        &+\iu \m{u}{-}{2}{3}\f{+}{\theta}\f{-}{\theta'}^*\Ikp{n_\mathrm{i}-1}{n}\Ik{n_\mathrm{f}}{n-1} \\
        &+ \m{u}{-}{2}{4}\f{+}{\theta}\f{z}{\theta'}^*\Ikp{n_\mathrm{i}}{n}\Ik{n_\mathrm{f}}{n-1} \\
        &+ \m{u}{+}{2}{1}\f{-}{\theta}\f{z}{\theta'}^*\Ikp{n_\mathrm{i}-1}{n-1}\Ik{n_\mathrm{f}-1}{n} \\
        &-\iu \m{u}{+}{2}{2}\f{-}{\theta}\f{+}{\theta'}^*\Ikp{n_\mathrm{i}}{n-1}\Ik{n_\mathrm{f}-1}{n} \\
        &+\iu \m{u}{+}{2}{3}\f{z}{\theta}\f{z}{\theta'}^*\Ikp{n_\mathrm{i}-1}{n-1}\Ik{n_\mathrm{f}}{n} \\
        &+ \m{u}{+}{2}{4}\f{z}{\theta}\f{+}{\theta'}^*\Ikp{n_\mathrm{i}}{n-1}\Ik{n_\mathrm{f}}{n}\Bigg]\Bigg\}.\\
    \end{split}
\end{equation}
\normalsize

\section{\label{app:DoubleIntegration}Integration bounds in 2-photon pair annihilation}
\begin{figure}
\includegraphics[width=0.8\textwidth]{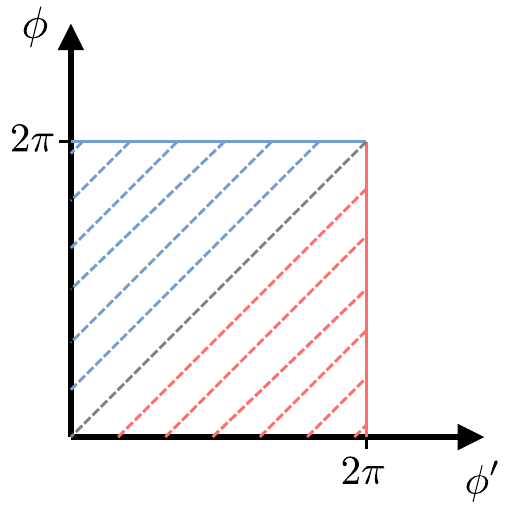}
\caption{The integration area is split into two triangles in the center-of-mass-type coordinates. The blue and red triangles correspond to negative and positive values of $\Delta\phi$, respectively. The dashed diagonal lines denote lines of constant $\Delta\phi$.} \label{fig:IntBounds}
\end{figure}
We want to evaluate an integral of the form
\begin{equation}
    I = \int_0^{2\pi}\dd{\phi}\int_0^{2\pi}\dd{\phi'}\abs{M_{2PA}}^2,
\end{equation}
where the squared matrix element is independent of ${\Phi = (\phi+ \phi')/2}$ and $2\pi$-periodic in $\Delta\phi = \phi'-\phi$. Thus, it is most convenient to perform the integration using the variables $\Phi$ and $\Delta\phi$. The Jacobian of this transformation is unity, but special care needs to taken to get the correct integration bounds. The integral in the new variables reads
\begin{equation}
    \begin{split}
    I &= \int_0^{2\pi}\dd{\Delta\phi} \int_\frac{\Delta\phi}{2}^{2\pi-\frac{\Delta\phi}{2}} \dd{\Phi}\abs{M_{2PA}}^2 \\
    &+ \int_{-2\pi}^0\dd{\Delta\phi} \int_{-\frac{\Delta\phi}{2}}^{2\pi+\frac{\Delta\phi}{2}} \dd{\Phi}\abs{M_{2PA}}^2 \\
    &= \int_0^{2\pi}\dd{\Delta\phi} \left(2\pi - \Delta\phi\right)\abs{M_{2PA}}^2 \\
    &+ \int_{-2\pi}^0\dd{\Delta\phi} \left(2\pi + \Delta\phi\right)\abs{M_{2PA}}^2 \\
    &= \int_0^{2\pi}\dd{\Delta\phi} \left(2\pi - \Delta\phi\right)\abs{M_{2PA}}^2 + \int_{0}^{2\pi}\dd{\Delta\phi}\Delta\phi\abs{M_{2PA}}^2 \\
    &= 2\pi\int_0^{2\pi}\dd{\Delta\phi}\abs{M_{2PA}}^2,
    \end{split}
\end{equation}
where we have made use of the $2\pi$-periodicity of the squared matrix element in $\Delta\phi$. The two different integration bounds correspond to integrating the lower-right and upper-left triangles of the full $2\pi\cross2\pi$ integration area separately, as shown in Fig. \ref{fig:IntBounds}.

\end{document}